\documentclass[a4paper,10pt,openany]{article}
\usepackage{setspace}
\usepackage{hyperref}
\usepackage{pst-3d}
\usepackage{pst-3dplot}
\usepackage{etex}
\usepackage[utf8]{inputenc}
\usepackage[english]{babel}
\usepackage{amsmath,amssymb,amsthm,mathrsfs,amsfonts,dsfont}
\usepackage{graphicx}
\usepackage[small, skip=10pt]{caption}
\usepackage{epigraph}
\usepackage{indentfirst}
\usepackage{booktabs}
\usepackage{fancyhdr}
\usepackage{vmargin}
\usepackage{feynmf}
\usepackage{yfonts}
\usepackage{lscape}
\usepackage{subfig}
\usepackage{textcomp}
\usepackage{pstricks}
\usepackage[metapost]{mfpic}
\usepackage{fancybox}
\usepackage{slashed}
\usepackage[verbose]{wrapfig}
\usepackage{pifont}
\usepackage{keystroke}
\usepackage{appendix}
\usepackage{enumitem}
\usepackage{array}
\usepackage{colortbl}
\usepackage{alltt}
\usepackage{boxedminipage}
\usepackage{tikz}
\usepackage{calc}
\usepackage{subfloat}
\usepackage{multirow}
\usepackage{cmll}
\usepackage{multicol}
\definecolor{color1}{RGB}{204,0,51}
\definecolor{color2}{RGB}{159,182,205}
\usepackage{bookmark}

\usepackage{verbatim}
\usepackage[vcentermath]{youngtab}
\usepackage{young}
\usepackage{pst-grad} 
\usepackage{pst-plot}
\usepackage{transparent}
\usepackage{color,soul}
\usepackage{cancel}
\usepackage{bm}
\usepackage{pifont}
\usepackage{placeins}
\usepackage{mathtools}
\usepackage{rotating}
\usepackage[refpage]{nomencl}
\usepackage{hhline}
\usepackage{marginnote}
\usepackage{upgreek}
\usepackage{stackengine}
\usepackage{ytableau}
\usepackage{listings}
\usepackage{calligra}
\usepackage{afterpage}
\usepackage{cite}
\usepackage{cleveref}
\usepackage{makecell}
\usepackage{accents}
\usepackage{breqn}
\usepackage{environ}

\usepackage{tikz-cd}
\usetikzlibrary{cd}

\usepackage[final]{showlabels}

\def\rme{\mathrm{e}}
\def\rmd{\mathrm{d}}
\def\rmR{\mathrm{R}}
\def\mcR{\mathcal{R}}
\newcommand{\lr}{\left(}
\newcommand{\rr}{\right)}

\definecolor{darkergreen}{rgb}{0.0, 0.5, 0.0}

\allowdisplaybreaks

\sethlcolor{yellow}
\definecolor{boh}{RGB}{79,47,79}

\hypersetup{
linkbordercolor={red}
}

\hypersetup{
    unicode=false,          
    pdftoolbar=true,        
    pdfmenubar=true,        
    pdffitwindow=false,     
    pdfstartview={FitH},    
    pdftitle={A Non-relativistic Limit of NS-NS Gravity},    
    pdfauthor={Eric Bergshoeff, Johannes Lahnsteiner, Luca Romano, Jan Rosseel, Ceyda Simsek},     
    pdfsubject={},   
    pdfnewwindow=true,      
    colorlinks=false,       
    linkcolor=red,          
    citecolor=red,        
    filecolor=red,      
    urlcolor=red           
    linkbordercolor={red},
    citebordercolor={red},
    urlbordercolor={red}
}

\makeatletter

\newcommand{\Rmnum}[1]{\expandafter\@slowromancap\romannumeral #1@}
\makeatother

\usepackage{alphalph}

\makeatletter
\newalphalph{\aalphalph}[mult]{\alphalph@alph}{26}
\newcommand{\alphalphval}[1]{%
  \@ifundefined{c@#1}{
    \aalphalph{#1}
  }{%
    \aalphalph{\value{#1}}
  }
}
\makeatother

\usepackage{color}

\def\chapterautorefname~#1\null{Chap.~(#1)\null}
\def\sectionautorefname~#1\null{Sec.~(#1)\null}
\def\subsectionautorefname~#1\null{sub--Sec.~(#1)\null}
\def\figureautorefname~#1\null{Fig.~(#1)\null}
\def\tableautorefname~#1\null{Tab.~(#1)\null}
\def\equationautorefname~#1\null{eq.~(#1)\null}

\def\equationautorefname~#1\null{eq.~(#1)\null}

\NewEnviron{multieq}[1][2]{

\begin{multicols}{#1}
\begin{subequations}
\setlength{\abovedisplayskip}{-15pt}
\allowdisplaybreaks
\begin{align}
\BODY
\end{align}
\end{subequations}
\end{multicols}
\setlength{\parindent}{0pt}
}

\DeclareMathAlphabet\mathbfcal{OMS}{cmsy}{b}{n}

\usepackage[yyyymmdd]{datetime}

\title{\bf A Non-Relativistic Limit of NS-NS Gravity}
\date{}

\begin{document}

\begin{flushright}
\small
\normalsize
\end{flushright}
{\let\newpage\relax\maketitle}
\maketitle
\def\equationautorefname~#1\null{eq.~(#1)\null}
\def\tableautorefname~#1\null{tab.~(#1)\null}

\vspace{0.8cm}

\begin{center}


\renewcommand{\thefootnote}{\alph{footnote}}
{\sl\large E.~A.~Bergshoeff$^{~1}$}\footnote{Email: {\tt e.a.bergshoeff[at]rug.nl}},
{\sl\large J.~Lahnsteiner$^{~1}$}\footnote{Email: {\tt j.m.lahnsteiner[at]outlook.com}},
{\sl\large L.~Romano$^{~1}$}\footnote{Email: {\tt lucaromano2607[at]gmail.com}},\\[.1truecm]
{\sl\large J.~Rosseel$^{~2}$}\footnote{Email: {\tt jan.rosseel[at]univie.ac.at}} and
{\sl\large C.~\c Sim\c sek$^{~1}$}\footnote{Email: {\tt c.simsek[at]rug.nl}}

\setcounter{footnote}{0}
\renewcommand{\thefootnote}{\arabic{footnote}}

\vspace{0.5cm}

${}^1${\it Van Swinderen Institute, University of Groningen\\
Nijenborgh 4, 9747 AG Groningen, The Netherlands}\\
\vskip .2truecm
${}^2${\it Faculty of Physics, University of Vienna,\\
Boltzmanngasse 5, A-1090, Vienna, Austria }\\

\vspace{1.8cm}


{\bf Abstract}
\end{center}
\begin{quotation}
  {\small
    We discuss a particular non-relativistic limit of NS-NS gravity that can be taken at the level of the action and equations of motion, without imposing any geometric constraints by hand. This relies on the fact that terms that diverge in the limit and that come from the Vielbein in the Einstein-Hilbert term and from the kinetic term of the Kalb-Ramond two-form field cancel against each other. This cancelling of divergences is the target space analogue of a similar cancellation that takes place at the level of the string sigma model between the Vielbein in the kinetic term and the Kalb-Ramond field in the Wess-Zumino term. The limit of the equations of motion leads to one equation more than the limit of the action, due to the emergence of a local target space scale invariance in the limit. Some of the equations of motion can be solved by scale invariant geometric constraints. These constraints define a so-called Dilatation invariant  String Newton-Cartan geometry.
}
\end{quotation}

\newpage

\tableofcontents

\section{Introduction}\label{sec:intro}

Non-relativistic (NR) string theory in flat space-time has been proposed a long time ago \cite{Gomis:2000bd,Danielsson:2000gi}. The generalization from a flat to special curved backgrounds was considered a few years later \cite{Gomis:2005pg}. Closed bosonic NR string theory in general curved backgrounds, on the other hand, has been constructed only recently. This has been done either by taking a NR limit \cite{Bergshoeff:2018yvt,Bergshoeff:2019pij} or by null reduction \cite{Harmark:2017rpg,Harmark:2018cdl,Harmark:2019upf} of the worldsheet action for a relativistic string in a generic background.\,\footnote{For other recent work on NR strings in a curved background, see
\cite{Kluson:2018egd,Kluson:2018vfd,Kluson:2019ifd,Roychowdhury:2019qmp}.}
This work showed that the natural target space geometry of the NR string theory of \cite{Gomis:2000bd,Danielsson:2000gi} in arbitrary backgrounds, is given by a NR Newton-Cartan-like geometry with co-dimension two foliation that is referred to as String Newton-Cartan (SNC) geometry \cite{Andringa:2012uz}.\,\footnote{For earlier work on SNC geometry, see \cite{Gomis:2005pg,Brugues:2004an,Brugues:2006yd}.}
The NR string then couples to the background fields of SNC geometry, as well as to a Kalb-Ramond (KR) and dilaton field. All these background fields must satisfy equations of motion that ensure (one loop) quantum Weyl invariance of the NR string worldsheet action \cite{Gomis:2019zyu,Yan:2019xsf}.\,\footnote{See also \cite{Gallegos:2019icg}.} The case of a NR open string in a curved background has been discussed recently in \cite{Gomis:2020fui,Gomis:2020izd}.

SNC geometry is a particular case of what can be called `$p$-brane Newton-Cartan geometry'. The latter term refers to $D$-dimensional Newton-Cartan-like geometries that can be written as a co-dimension $p+1$ foliation. These geometries are then equipped with two degenerate metrics, one of rank $D-p-1$ on the leaves of the foliation and one of rank $p+1$ on the foliation's co-dimension $p+1$ part. The directions spanned by the co-dimension $p+1$ part can be viewed as lying along a NR $p$-brane worldvolume, while the leaves of the foliation represent directions that are transversal to this worldvolume.

Such $p$-brane geometries can be obtained as a `$p$-brane NR limit' of the Lorentzian geometry that underlies General Relativity. This limit can be conveniently discussed in the Vielbein formulation of Lorentzian geometry. To do this, one considers the relativistic Vielbein $E_\mu{}^{\hat{A}}$ and splits the $D$-dimensional flat SO$(1,D-1)$ index $\hat{A}$ into a flat `worldvolume' index $A=0,\cdots,p$ and a flat `transversal' index $A^\prime = p+1,\cdots, D$. One then  redefines $E_\mu{}^{\hat{A}}$ as follows:
\begin{equation}\label{expansion}
 {E}_\mu{}^{A} = c\tau_\mu{}^A +\frac{1}{c}m_\mu{}^A\,,\hskip 1.5truecm {E}_\mu{}^{A'} = e_\mu{}^{A'}\,,
\end{equation}
where $c$ is a contraction parameter.
As it stands, this redefinition is not invertible. To make it invertible one typically introduces (and redefines) a $p+1$-form gauge field by hand. In the case of string theory, i.e.~$p=1$, the role of this 2-form gauge field is played by  the KR 2-form field.
The NR limit of a quantity, constructed out of $E_\mu{}^{\hat{A}}$, is then obtained by plugging in this redefinition in the object of interest, expanding the result in powers of $c^{-2}$ and formally taking the limit $c\rightarrow \infty$, by retaining only the leading order in this expansion.\,\footnote{Strictly speaking, when taking the limit that $c$ goes to infinity, we mean that one first redefines $c \rightarrow \lambda c$ and then takes the limit where the dimensionless contraction parameter $\lambda$ goes to infinity.} By doing this for the Lorentzian metric, one obtains the degenerate metric, with `longitudinal Vielbein' $\tau_\mu{}^A$, on the co-dimension $p+1$ part of a $p$-brane Newton-Cartan geometry, while the inverse Lorentzian metric leads to the degenerate metric, with `transversal Vielbein' $e_\mu{}^{A^\prime}$ on the foliation leaves. The SO$(1,D-1)$ local Lorentz transformations get contracted in the limit to `homogeneous $p$-brane Galilei symmetries', consisting of worldvolume Lorentz transformations, transversal rotations and Galilean-type boosts between transversal and worldvolume directions. These homogeneous $p$-brane Galilei symmetries are part of a larger symmetry group that includes translations. The field $m_\mu{}^A$ can then be identified as a gauge field for a non-central (central in case $p=0$) extension of this larger symmetry group.

The NR limit can also be taken for the relativistic spin connection ${\Omega}_\mu{}^{\hat A\hat B}$, to obtain NR spin connections for the above mentioned homogeneous $p$-brane Galilei symmetries. Plugging the redefinitions \eqref{expansion} in the spin connection components ${\Omega}_\mu{}^{\hat A\hat B}$ in the second-order formulation\,\footnote{Here, we consider the second-order formulation of General Relativity, in anticipation of the extension of the results of this paper to NS-NS supergravity, for which no first-order formulation is available in the literature. The expression for the relativistic dependent spin-connection $\Omega_\mu{}^{\hat{A} \hat{B}}$ can be found in Appendix \ref{ssec:LorGeom}.} and expanding the result in powers of $c^{-2}$, one finds that the leading order terms of this expansion do not transform as NR spin connections for the homogeneous $p$-brane Galilei symmetries. Instead, it is the subleading order terms of this expansion that give rise to proper NR spin connections. For this reason, the leading order terms are considered to be divergent. In order to obtain correct NR spin connections in the NR limit, one then requires that these divergent terms vanish. This can be done by imposing the following `zero torsion constraint' on the longitudinal Vielbein $\tau_\mu{}^A$:
\begin{equation}\label{stringconstraint}
D_{[\mu}(\omega)\tau_{\nu]}{}^A=0\,,
\end{equation}
where the derivative $D_\mu(\omega)$ is covariantized with respect to longitudinal Lorentz transformations, using a spin connection $\omega_\mu{}^{AB}$. Part of this constraint \eqref{stringconstraint} is identically satisfied once the dependent expression for $\omega_\mu{}^{AB}$ in terms of the $\tau_\mu{}^A$ and their projective inverses $\tau_A{}^\mu$,\,\footnote{For an explicit expression of $\omega_\mu{}^{AB}$ in terms of $\tau_\mu{}^A$ and its inverse, see Appendix \ref{sec:stringGalileigeometry}.} is plugged in. However, not all of \eqref{stringconstraint} is identically satisfied in this way and \eqref{stringconstraint} thus leads to a genuine constraint on $\tau_\mu{}^A$ and the geometry. In case $p=0$ for instance, this geometric constraint entails that the time-like Vielbein of Newton-Cartan geometry is closed, implying that the space-time admits an absolute time direction. For the string case ($p=1$), the constraints \eqref{SNCconstraints} define the torsionless SNC geometry that was considered as target space-time for the NR string in \cite{Bergshoeff:2018yvt,Bergshoeff:2019pij}. Imposing the zero torsion constraint, it was shown in \cite{Bergshoeff:2015sic} that the 0-brane limit of the equations of motion of General Relativity leads to the equations of motion of Newton-Cartan gravity. Similarly, it was shown that the target space equations of motion for the NR string theory of \cite{Bergshoeff:2018yvt,Bergshoeff:2019pij} arise from the 1-brane limit of the equations of motion of NS-NS gravity, upon imposition of \eqref{stringconstraint}. Note that in both cases, the constraint \eqref{stringconstraint} was imposed by hand and can not be considered as an equation of motion that follows from a NR action. For that reason, the limit was in both cases taken at the level of the equations of motion and not at the level of the action.

When taking the NR limit of supersymmetric theories, imposing the constraint \eqref{stringconstraint} (or similar geometric constraints) by hand can be problematic. Indeed, when imposing constraints by hand, one also needs to impose their supersymmetry variations as constraints in order to maintain supersymmetry. This typically leads to a tower of constraints on top of the equations of motion of the NR theory. There is then a danger that the NR limit does not lead to the most general possible theory or even leads to an overconstrained theory. When taking the NR limit of supergravity theories, it would thus be better, if one were able to take the NR limit of the action or equations of motion, without imposing the zero torsion constraint as an a priori constraint. Some of the NR equations of motion might then take the form of differential or algebraic constraints for the components of $D_{[\mu}(\omega)\tau_{\nu]}{}^A$. These, however, do not give rise to a tower of extra constraints on top of the NR equations of motion, since they correspond to NR equations of motion themselves. If the NR limit is taken consistently, their supersymmetry variations should thus also give rise to equations of motion.

In this paper, we will focus exclusively on the $p=1$ case of SNC geometry. Motivated by supersymmetry, we then address the question whether it is possible to take the NR limit of relativistic gravity, without imposing the zero torsion constraint \eqref{stringconstraint} by hand. In a similar spirit as in the work of \cite{VandenBleeken:2017rij}, we will show that this is possible for the matter-coupled relativistic gravity theory that corresponds to NS-NS gravity. We will see that in order to achieve this, we not only have to adopt the redefinitions \eqref{expansion}, but we similarly have to expand the NS-NS two-form field $B_{\mu\nu}$ as
\begin{equation} \label{Bexpansionintro}
  B_{\mu\nu} = -c^2 \tau_\mu{}^A \tau_\nu{}^B \epsilon_{AB} + b_{\mu\nu} \,,
\end{equation}
where $b_{\mu\nu}$ corresponds to the two-form field of the NR theory that results from taking the $c\rightarrow \infty$ limit. The redefinitions \eqref{expansion} and \eqref{Bexpansionintro} are then the same as the ones used to obtain the NR string worldsheet action as the NR limit of the relativistic string action. In that case, a fine-tuning between the Vielbein and the NS-NS two-form field leads to a cancellation of divergences when taking the NR limit of the relativistic string action, so that a non-trivial NR action is obtained. We will show here that a similar mechanism takes place when taking the NR limit of NS-NS gravity, so that a non-trivial NR theory is obtained without imposing any geometric constraints by hand. Since no constraints need to be imposed in the process, the NR limit can be taken both for the equations of motion and the action of NS-NS gravity. We then find that both the action and equations of motion of the resulting NR theory exhibit an emerging local scale invariance, under which the longitudinal Vielbein $\tau_\mu{}^A$ scales non-trivially.

The appearance of this emerging local scale symmetry has two consequences. First, it implies that the NR action leads to one equation of motion less than its relativistic counterpart. It turns out that the missing equation of motion is important, as it corresponds to the analog of the Poisson equation of NR gravity. We will see that this equation of motion is recovered by considering the NR limit of the equations of motion of NS-NS gravity, so that taking the NR limit of the action is not equivalent to taking the NR limit of the equations of motion. Secondly, since the zero torsion constraint \eqref{stringconstraint} is not invariant under local scale symmetries, it can not arise as one of the NR equations of motion. We will indeed see that some of the NR equations of motion that we find amount to algebraic and differential equations for $\tau_\mu{}^A$ allowing as a solution  the following set of constraints that is weaker than the SNC geometric constraints given in \eqref{stringconstraint}:\footnote{The second constraint is sufficient to define a globally well-defined co-dimension two foliation, called integrable distribution \cite{Frankel:1997ec}.}
\begin{align} \label{TTSintro}
  e_{A^\prime}{}^\mu \tau_{\{A|}{}^\nu \partial_{[\mu} \tau_{\nu]| B\}} =0\,,\qquad\mathrm{and}\qquad e_{A^\prime}{}^\mu e_{B^\prime}{}^\nu \partial_{[\mu} \tau_{\nu]}{}^A =0\,.
\end{align}
Here, $\tau_A{}^\mu$, $e_{A^\prime}{}^\mu$ are (projective) inverses of $\tau_\mu{}^A$,$e_\mu{}^{A^\prime}$ and $\{AB\}$ indicates the symmetric traceless part of $AB$. This set of constraints is invariant under the local scale symmetry of the NR theory. Compared to the SNC constraints \eqref{stringconstraint}, we have that
\begin{equation}
b_{A'} \equiv e_{A^\prime}{}^\mu \tau_{A}{}^\nu \partial_{[\mu} \tau_{\nu]}{}^A\,,
\end{equation}
which acts like the (transverse components of the) gauge field of the local scale symmetry, is non-zero. \footnote{We call the geometry defined by \eqref{stringconstraint} a String Newton-Cartan (SNC) geometry and \eqref{TTSintro} a Dilatation invariant String  Newton-Cartan (DSNC) geometry. The geometry without constraints will be referred to as a Torsional String Newton-Cartan (TSNC) geometry. Note that the  geometry defined by the second constraint only in  \eqref{TTSintro} is a string version of the Twistless Torsional Newton-Cartan (TTNC) geometry, found in Lifshitz holography \cite{Christensen:2013lma}.}

This paper is organized as follows. In section 2 we review the NR string worldsheet action and how it can be obtained from the NR limit to motivate the way we define the NR limit of NS-NS gravity. In the next section 3, we discuss how the NR limit can be taken for the NS-NS gravity action and give the equations of motion that stem from the resulting NR action.  In section 4, we discuss the NR limit of the equations of motion of NS-NS gravity and show how this gives rise to an extra Poisson equation that can not be derived from the NR action. In section 5, we compare our results to the target space equations of motion that were obtained by calculating the beta functions of NR string theory in SNC geometry, with the zero torsion constraint imposed \cite{Gomis:2019zyu,Yan:2019xsf}. Finally, we present our conclusions and discuss several applications and generalizations of our results. There are three appendices. In appendix A, we give our notation and conventions. Appendix B discusses details of the NR geometry that appears in our NR action and equations of motion. Appendix C contains details on how the NR limit of the action and equations of motion of NS-NS gravity is taken.

\section{The NR Bosonic String Action} \label{sec:NRstringaction}

The worldsheet action for the NR bosonic string in a generic background was derived in \cite{Bergshoeff:2018yvt,Bergshoeff:2019pij}, by taking a NR limit of the relativistic Polyakov string action, coupled to an arbitrary target space background. This leads to the following NR string action in the Polyakov form:
\begin{align}\label{eq:nraction}
 S_{\rm P} &= -\frac{T}{2}\int \rmd^2\sigma \, \left [
\sqrt{-h} h^{\alpha\beta}\partial_\alpha x^\mu \partial_\beta x^\nu H_{\mu\nu} + \epsilon^{\alpha\beta} \left(\lambda e_\alpha \tau_\mu + \bar{\lambda} \bar{e}_\alpha \bar{\tau}_\mu \right)\partial_\beta x^\mu \right]\nonumber \\ & \qquad -\frac{T}{2} \int \rmd^2 \sigma \, \epsilon^{\alpha\beta}\partial_\alpha x^\mu \partial_\beta x^\nu b_{\mu\nu}  + \frac{1}{4\pi} \int \rmd^2 \sigma \sqrt{-h} \, R^{(2)}(h)\, \left(\phi- \tfrac{1}{4} \ln G \right)\,.
\end{align}
Here $T$ is the string tension, $\sigma^\alpha\ (\alpha=0,1)$ are the worldsheet coordinates and $x^\mu (\sigma)$, $\mu = 0, 1, \cdots, 9$, are the string embedding coordinates. We have denoted the worldsheet metric, its determinant and its Ricci scalar by $h_{\alpha\beta}$, $h$ and $R^{(2)}(h)$ respectively. We have furthermore introduced a Zweibein $e_\alpha{}^a$ ($a=0,1$) for $h_{\alpha\beta}$ via $h_{\alpha\beta} = e_\alpha{}^a e_\beta{}^b \eta_{ab}$ (with $\eta_{ab} = \mathrm{diag}(-1,1)$) and denoted its components in a light-cone basis as
\begin{equation}
  \label{eq:zweibeinlc}
  e_\alpha = e_\alpha{}^0 + e_\alpha{}^1 \,, \qquad \qquad \bar{e}_\alpha = e_\alpha{}^0 - e_\alpha{}^1 \,.
\end{equation}
The second term in \eqref{eq:nraction} includes two extra worldsheet fields $\lambda(\sigma)$, $\bar{\lambda}(\sigma)$ that appear as Lagrange multipliers. We refer to \cite{Bergshoeff:2019pij} for details on how these fields appear in the NR limit.

The NR string couples in the action \eqref{eq:nraction} to background fields that we take to be ten-dimensional ones.\footnote{Strictly speaking, the background fields for the critical NR bosonic string are 26-dimensional. Here however, we consider ten-dimensional backgrounds, since we have the superstring in mind. We thus view \eqref{eq:nraction} as the bosonic part of a NR superstring action.} They are given by
\begin{equation}
\{\tau_\mu{}^A\,, e_\mu{}^{A^\prime}\,, m_\mu{}^A\,, b_{\mu\nu}\,, \phi\}\,,\qquad \qquad A=0,1\,;\ \ A^\prime = 2,\cdots, 9\,,
\end{equation}
representing the longitudinal Vielbein $\tau_\mu{}^A$, the transverse Vielbein $e_\mu{}^{A^\prime}$ and the non-central charge 
gauge field $m_\mu{}^A$ of SNC geometry, as well as the KR field $b_{\mu\nu}$ and the dilaton $\phi$.
The first term in the action \eqref{eq:nraction} is the kinetic term and contains the so-called `transverse metric' $H_{\mu\nu}$ that is given in terms of the SNC background fields by\,\footnote{Note that this metric is strictly speaking only transverse in the absence of the second term.}
\begin{equation}\label{definition}
	H_{\mu\nu} = e_\mu{}^{A'} e_\nu{}^{B'} \delta_{A'B'} + \bigl( \tau_\mu{}^A m_\nu{}^B + \tau_\nu{}^A m_\mu{}^B \bigr) \, \eta_{AB}\,.
      \end{equation}
The fields $\tau_\mu$, $\bar{\tau}_\mu$ in the second term correspond to $\tau_\mu{}^A$ in a light-cone basis:
      \begin{equation}
        \label{eq:deftautaubar}
        \tau_\mu = \tau_\mu{}^0 + \tau_\mu{}^1 \,, \qquad \qquad \bar{\tau}_\mu = \tau_\mu{}^0 - \tau_\mu{}^1 \,.
      \end{equation}
The third term in the action \eqref{eq:nraction} describes the Wess-Zumino coupling of the background KR field $b_{\mu\nu}$ to the string. Furthermore, the object $G$ in the last term of \eqref{eq:nraction} was defined in \cite{Bergshoeff:2019pij} as the limit of the metric determinant 
    \begin{align}\label{eq:metricdet}
        G=e^2\equiv-\lim_{c\to\infty}( c^{-4}\det G_{\mu\nu})\,,\quad\mathrm{where}\quad e = \det\big(\tau^A, e^{A'}\big) \equiv \varepsilon^{\mu_1\cdots \mu_{10}}\,\tau_{\mu_1}{}^0\tau_{\mu_2}{}^1 e_{\mu_2}{}^2\cdots e_{\mu_{10}}{}^9\,.
    \end{align}One can also derive a NR string action in Nambu-Goto form, by integrating out the Lagrange multipliers $\lambda$, $\bar{\lambda}$ \cite{Bergshoeff:2018yvt}. The equations of motion of $\lambda$, $\bar{\lambda}$ correspond to the constraints
\begin{equation}
  \label{eq:constraintsllbar}
  \epsilon^{\alpha\beta} e_\alpha \tau_\mu \partial_\beta x^\mu = 0 \,, \qquad \qquad \epsilon^{\alpha\beta} \bar{e}_\alpha \bar{\tau}_\mu \partial_\beta x^\mu = 0 \,.
\end{equation}
These constraints are solved by
\begin{equation}
  \label{eq:solconstraintsllbar}
  h_{\alpha\beta} = \alpha(x) \tau_{\alpha \beta} \,,
\end{equation}
where $\alpha(x)$ is an arbitrary proportionality factor and
\begin{equation}
  \label{eq:deftaumetric}
  \tau_{\alpha\beta} \equiv \tau_\mu{}^A \tau_{\nu}{}^B \eta_{AB} \partial_\alpha x^\mu \partial_\beta x^\nu \,.
\end{equation}
Plugging the solution \eqref{eq:solconstraintsllbar} in the NR Polyakov action \eqref{eq:nraction}, leads to the NR Nambu-Goto action, given by
\begin{equation}
  \label{eq:nractionNG}
 S_{\rm NG} = -\frac{T}{2}\int \rmd^2\sigma \, \left [
\sqrt{-\text{det}(\tau_{\gamma \delta})}\tau^{\alpha\beta}\partial_\alpha x^\mu \partial_\beta x^\nu H_{\mu\nu}
+\epsilon^{\alpha\beta}\partial_\alpha x^\mu \partial_\beta x^\nu b_{\mu\nu}\right ] + S_{\rm dilaton}\,,
\end{equation}
where $S_{\rm dilaton}$ is the last term of \eqref{eq:nraction} (with $h_{\alpha\beta}$ replaced by the solution \eqref{eq:solconstraintsllbar}) and $\tau^{\alpha\beta}$ is the inverse of $\tau_{\alpha\beta}$. Ignoring $S_{\rm dilaton}$, this action can be obtained from a NR limit of the relativistic string action in Nambu-Goto form \cite{Harmark:2019upf}:
\begin{equation}
  \label{eq:SNGrel}
  S_{\mathrm{rel-NG}} = - T \int \rmd^2 \sigma \, \sqrt{-\mathrm{det}\left(E_\mu{}^{\hat{A}} E_{\nu \hat{A}} \partial_\alpha x^\mu \partial_\beta x^\nu \right)} - \frac{T}{2} \int \rmd^2 \sigma \, \epsilon^{\alpha\beta} \partial_\alpha x^\mu \partial_\beta x^\nu B_{\mu\nu} \,,
\end{equation}
where $E_\mu{}^{\hat{A}}$ is the relativistic ten-dimensional Vielbein and $B_{\mu\nu}$ the relativistic KR field. Splitting the index $\hat{A}$ in $A = 0,1$ and $A^\prime = 2, \cdots, 9$, the first two terms of \eqref{eq:nractionNG} are then obtained by plugging the following redefinitions (see \cite{Klebanov:2000pp,Danielsson:2000gi,Gomis:2000bd,Harmark:2019upf,Bergshoeff:2019pij} for early and recent references)
  \begin{align} \label{eq:redefsm}
    E_\mu{}^A &= c \tau_\mu{}^A + \frac{1}{c} m_\mu{}^A \,, \qquad \qquad E_\mu{}^{A^\prime} = e_\mu{}^{A^\prime} \,, \qquad \qquad
    B_{\mu\nu} = -c^2 \tau_\mu{}^A \tau_\nu{}^B \epsilon_{AB} + b_{\mu\nu} \,,
  \end{align}
in \eqref{eq:SNGrel} and taking the limit $c\rightarrow \infty$. The first two terms of \eqref{eq:nractionNG} constitute the terms at $\mathcal{O}(c^0)$ in an expansion of \eqref{eq:SNGrel} in powers of $c^{-2}$ (after the redefinitions \eqref{eq:redefsm} have been performed). When expanding, both terms in \eqref{eq:SNGrel} lead to a contribution at $\mathcal{O}(c^2)$ that diverges in the $c \rightarrow \infty$ limit. The divergent contribution that comes from the second, Wess-Zumino term, of \eqref{eq:SNGrel} however exactly cancels the contribution coming from the first, kinetic term, so that the $c \rightarrow \infty$ limit is well-defined.

The actions \eqref{eq:nraction}, \eqref{eq:nractionNG} are invariant under an abelian two-form symmetry, with parameters $\theta_\mu$, of the KR field
\begin{align}
  \delta b_{\mu\nu} = 2 \partial_{[\mu} \theta_{\nu]} \,,
\end{align}
as well as under local transformations of the background fields that we will refer to as `String Galilei symmetries' in this
paper.\,\footnote{Invariance of \eqref{eq:nraction} under String Galilei symmetries requires that one also assigns a non-trivial SO$(1,1)$-transformation rule to the Lagrange multipliers $\lambda$, $\bar{\lambda}$. Similar remarks hold for the Stueckelberg invariance. We refer to \cite{Bergshoeff:2019pij} for the details.}
These consist of longitudinal SO$(1,1)$ Lorentz transformations with parameter $\lambda_M$, transversal SO$(8)$ rotations with parameters $\lambda^{A^\prime B^\prime}$ and Galilean boosts with parameters $\lambda^{AA^\prime}$ and their non-trivial transformation rules are given by
\begin{align}
  \label{eq:stringgalsymmsm}
  \delta \tau_\mu{}^A &= \lambda_M \epsilon^{A}{}_B \tau_\mu{}^B \,, \qquad \qquad \qquad \qquad \delta e_\mu{}^{A^\prime} = \lambda^{A^\prime}{}_{B^\prime} e_\mu{}^{B^\prime} - \lambda_A{}^{A^\prime} \tau_\mu{}^A \,, \nonumber \\
  \delta m_\mu{}^A &= \lambda_M \epsilon^{A}{}_B m_\mu{}^B + \lambda^{A}{}_{A^\prime} e_\mu{}^{A^\prime} \,.
\end{align}
Note that $H_{\mu\nu}$ is invariant under these symmetries, so that the String Galilei invariance of the actions \eqref{eq:nraction}, \eqref{eq:nractionNG} is manifestly realized.

The actions \eqref{eq:nraction}, \eqref{eq:nractionNG} are furthermore also invariant under the following Stueckelberg symmetry 
 with parameters $c_\mu{}^A$,
given by 
\begin{equation}\label{oldstuckelberg}
    \delta b_{\mu\nu}= (\,c_{\mu}{}^A\tau_{\nu}^B- c_{\nu}{}^A\tau_{\mu}{}^B\,)\,\epsilon_{AB}\,,\hskip 1.5truecm
   \delta m_\mu{}^A = - c_\mu{}^A\,.
 \end{equation}
This symmetry is a direct consequence of the fact that we have introduced more non-relativistic than relativistic fields in \eqref{eq:redefsm}. This over-parametrization leads to the emergence of the shift symmetry. Note that invariance under \eqref{oldstuckelberg} is not manifest: it is only due to a non-trivial cancellation of the symmetry variation of the kinetic term with that of the Wess-Zumino term. The Stueckelberg symmetry \eqref{oldstuckelberg} is a reducible symmetry in the sense that the transformation rule of $b_{\mu\nu}$ and the transformation rule of $H_{\mu\nu}$, as induced by $\delta m_\mu{}^A = - c_\mu{}^A$, are formally invariant under a gauge symmetry, with singlet parameter $c$, given by
  \begin{equation}
  \delta c_\mu{}^A = \epsilon^{AB}\tau_{\mu B}\,c\,.
  \end{equation}
  The Stueckelberg symmetry is thus parametrized by only 19 independent parameters.

One can rewrite the action \eqref{eq:nractionNG} in a manifestly Stueckelberg invariant way, by moving the $m_\mu{}^A$ terms, that are part of the definition of $H_{\mu\nu}$, from the kinetic term of \eqref{eq:nractionNG} to the Wess-Zumino term, where they form a Stueckelberg-invariant combination, \begin{equation}\label{combination}
b_{\mu\nu} + ( m_{\mu}{}^A\tau_{\nu}^B- m_{\nu}{}^A\tau_{\mu}{}^B\,)\,\epsilon_{AB}\,.
\end{equation}
Equivalently, one can also fix the Stueckelberg symmetry by imposing the gauge-fixing condition
\begin{equation} \label{eq:mzerogf}
m_\mu{}^A=0 \,,
\end{equation}
after which the string action \eqref{eq:nractionNG} reads as follows:
\begin{align}\label{curvedindices}
 S_{\rm NGf} = -\frac{T}{2}\int \rmd^2\sigma \,  \left [
\sqrt{-\mathrm{det}(\tau_{\gamma\delta})}\tau^{\alpha\beta} e_\alpha{}^{A^\prime} e_\beta{}^{B^\prime}\delta_{A^\prime B^\prime}
+\epsilon^{\alpha\beta}\partial_\alpha x^\mu \partial_\beta x^\nu b_{\mu\nu}\right ] + S_{\rm dilaton}\,.
\end{align}
Note that in contrast to the actions \eqref{eq:nraction} and \eqref{eq:nractionNG}, the Stueckelberg gauge-fixed action \eqref{curvedindices} exhibits the Galilean boost symmetry in a non-manifest way that involves the KR field in a non-trivial manner. Indeed, the price one pays for fixing the Stueckelberg symmetry is that the KR field transforms under compensating Galilean boosts. In the action \eqref{curvedindices}, $b_{\mu\nu}$ thus transforms under Galilean boosts as follows:
 \begin{equation} \label{eq:deltabgf}
 \delta b_{\mu\nu} = - 2\,\epsilon_{AB}\lambda^A{}_{A'}\tau_{[\mu}{}^B\,e_{\nu]}{}^{A'}\,.
\end{equation}
Checking boost invariance of \eqref{curvedindices} then requires cancelling a contribution from the boost variation of $e_\mu{}^{A^\prime}$ (given in \eqref{eq:stringgalsymmsm}) in the first term of \eqref{curvedindices} against a contribution from the variation \eqref{eq:deltabgf} of $b_{\mu\nu}$ in the second term.

It is worth pointing out that ordinarily, the longitudinal components of $m_\mu{}^A$ capture the information of NR gravity that is contained in the Newton potential \cite{Andringa:2012uz}. The effect of fixing the Stueckelberg symmetry, as in \eqref{eq:mzerogf}, is that the Newton potential is contained in the longitudinal component\footnote{We refer to Appendix \ref{ssec:indexconversions} for details on how curved indices are turned into flat ones in the different sections of this paper.} $b_{AB}$ of the KR field. This can be seen from the fact that the gauge-fixing \eqref{eq:mzerogf} is equivalent to replacing $b_{\mu\nu}$ by the Stueckelberg invariant combination \eqref{combination} that contains the field $m_\mu{}^A$. In the following sections of this paper, we will work with this Stueckelberg gauge-fixed formulation. We will therefore also refer to $b_{AB}$ as `the Newton potential'.

A final non-trivial property of the string action {\eqref{eq:nraction}} is that it has an emerging local dilatation symmetry, with parameter $\lambda_D$, given by\footnote{{In order to show invariance one also needs the transformation rule for the Stueckelberg field $\delta m_\mu{}^A = -\lambda_D m_\mu{}^A$, and the Lagrange multipliers $\delta \lambda = -\lambda_D \lambda$, $\delta \bar{\lambda} = -\lambda_D \bar{\lambda}$. All the other fields in \eqref{eq:nraction} have zero charge under local dilatations.}}
\begin{align}\label{localD}
  \delta \tau_\mu{}^A &= \lambda_D \tau_\mu{}^A\,, \qquad  \qquad \qquad \delta \phi = \lambda_D \,.
\end{align}
This symmetry was not present in the relativistic case. It arises due to the fact that the background fields couple to a string and not, for instance, to a membrane. It will play an important role in the remainder of this paper.

Let us finish this section by comparing the above discussion of SNC geometry as a NR limit with the one, given in previous work \cite{Bergshoeff:2019pij}. In \cite{Bergshoeff:2019pij}, the elements of SNC geometry were derived from a NR limit of General Relativity that closely reproduces a formulation of SNC geometry that is obtained from gaugings of underlying `String Bargmann or String Newton-Cartan' space-time symmetry algebras \cite{Andringa:2012uz}. These symmetry algebras contain a non-central extension $Z_A$, whose corresponding gauge field is given by $m_\mu{}^A$. This $Z_A$ symmetry was then argued to be a symmetry of the NR string action, in case the target space SNC geometry obeys the zero torsion constraint \eqref{stringconstraint} \cite{Bergshoeff:2019pij}. In this paper, we view SNC geometry as the target space geometry that arises when taking the NR limit of the bosonic string actions, as explained in this section.  When keeping the $m_\mu{}^A$ field, the NR limit of the string action leads to the Stueckelberg symmetries \eqref{oldstuckelberg} that can not be interpreted as symmetries of the String Bargmann or String Newton-Cartan algebras. When fixing $m_\mu{}^A = 0$, the $b_{\mu\nu}$ field, that can not be viewed as a gauge field of String Bargmann or String Newton-Cartan symmetries, transforms non-trivially under Galilean boosts and becomes part of the fields of SNC geometry. Since in either case, the resulting geometry can not be interpreted as a gauging of the String Bargmann or String Newton-Cartan algebra, there is no longer a reason to require the presence of the $Z_A$ symmetry. This $Z_A$ symmetry can be regained by imposing the zero torsion constraint \eqref{stringconstraint}, in which case it can be viewed as a special case of the Stuckelberg symmetries \cite{Harmark:2019upf}. In this paper, we do not wish to impose the zero torsion constraint and we will thus not necessarily have the $Z_A$ symmetry.

\section{The NR Limit of the NS-NS Gravity Action} \label{sec:limitaction}

In the previous section, we reviewed how the NR limit, defined in eqs. \eqref{eq:redefsm}, can be used to obtain the NR string worldsheet action \eqref{eq:nractionNG}. Here, we will show that this limit can also be applied in a well-defined way to the NS-NS gravity action to yield an action for all target space background fields of NR string theory. We will first discuss this NR limit in more detail and, in particular, show that it reproduces the correct transformation rules of the NR background fields under, e.g., String Galilei symmetries. After having recalled the relativistic NS-NS gravity action, we will then take its NR limit and discuss the resulting action.

\subsection{Preliminaries}\label{Preliminaries}

The field content of ten-dimensional relativistic NS-NS gravity is given by the dilaton $\Phi$, the KR two-form field $B_{\mu\nu}$ and the metric $G_{\mu\nu}$ that we will describe in terms of the Vielbein $E_\mu{}^{\hat{A}}$. To define the NR limit, we first redefine these fields, using a parameter $c$, as follows:
\begin{align}
    E_\mu{}^A = c\,\tau_\mu{}^A\,,\qquad E_\mu{}^{A'} = e_\mu{}^{A'}\,,\qquad B_{\mu\nu} = -c^2\,\epsilon_{AB}\,\tau_\mu{}^A\tau_\nu{}^B + b_{\mu\nu}\,,\qquad \Phi = \phi+\ln c\,.\label{eq:R=NR}
\end{align}
This corresponds to the redefinitions \eqref{eq:redefsm}, where we have however adopted the condition \eqref{eq:mzerogf} that fixes the Stueckelberg symmetries \eqref{oldstuckelberg} of the NR string worldsheet action. As shown by the use of these redefinitions in deriving the NR string worldsheet action \eqref{eq:nractionNG}, the fields $\tau_\mu{}^A$, $e_\mu{}^{A^\prime}$, $b_{\mu\nu}$ and $\phi$ correspond to the background fields of NR string theory, once the limit $c\rightarrow \infty$ has been taken.

Let us first discuss how eqs. \eqref{eq:R=NR} can be used to derive the transformation rules of $\tau_\mu{}^A$, $e_\mu{}^{A^\prime}$, $b_{\mu\nu}$ and $\phi$ under String Galilei symmetries from the transformations of $E_\mu{}^{\hat{A}}$, $B_{\mu\nu}$ and $\Phi$ under SO$(1,9)$ Lorentz transformations. To do this, we use the fact that the redefinitions \eqref{eq:R=NR} are invertible with inverse given by
\begin{align}
    \tau_\mu{}^A = c^{-1}\,E_\mu{}^A\,,\qquad e_\mu{}^{A'}=E_\mu{}^{A'}\,,\qquad b_{\mu\nu} = B_{\mu\nu} + \epsilon_{AB}\,E_\mu{}^A E_\nu{}^B\,,\qquad \phi = \Phi\,-\ln c\,.\label{eq:NR=R}
\end{align}
We can also introduce inverse Vielbeine $\tau_A{}^\mu=c\,E_A{}^\mu$ and $e_{A'}{}^\mu = E_{A'}{}^\mu$ that satisfy
\begin{align} \label{eq:invvielbeine}
  & \tau_A{}^\mu \tau_\mu{}^B = \delta_A^B \,, \qquad e_{A^\prime}{}^\mu e_\mu{}^{B^\prime} = \delta_{A^\prime}^{B^\prime} \,, \notag\\
  &\tau_A{}^\mu e_\mu{}^{A^\prime} = 0 \,, \qquad~~ e_{A^\prime}{}^\mu \tau_\mu{}^A = 0 \,,  \\
  & \tau_A{}^\mu\tau_\nu{}^A + e_{A'}{}^\mu e_\nu{}^{A'} = \delta^\mu_\nu\,.\nonumber
\end{align}
The transformation rules of $E_\mu{}^{\hat{A}}$, $B_{\mu\nu}$ and $\Phi$ under SO$(1,9)$ Lorentz transformations and the abelian two-form gauge symmetry of the KR field are given by
\begin{align} \label{eq:relrules}
        &\delta E_\mu{}^A = \Lambda_M \epsilon^A{}_B E_\mu{}^B + \Lambda^A{}_{A^\prime} E_\mu{}^{A^\prime} \,,&\delta E_\mu{}^{A^\prime} &= -\Lambda_A{}^{A^\prime} E_\mu{}^A + \Lambda^{A^\prime}{}_{B^\prime} E_\mu{}^{B^\prime} \,,\notag \\
        &\delta B_{\mu\nu} = 2\,\partial_{[\mu} \Theta_{\nu]} \,, &\delta \Phi &= 0 \,,
\end{align}
where $\Theta_\mu$ is the parameter of the two-form gauge symmetry and we have split SO$(1,9)$ into SO$(1,1)$ (with parameter $\Lambda^{AB} = \Lambda_M \epsilon^{AB}$), SO$(8)$ (with parameters $\Lambda^{A^\prime B^\prime}$) and the remaining boost transformations (with parameters $\Lambda^{A A^\prime}$).
Using the field redefinitions \eqref{eq:R=NR}, their inverses \eqref{eq:NR=R} and the following redefinitions of the symmetry parameters
\begin{align}
    \Lambda_M = \lambda_M \,,\qquad \Lambda^{AA'}{=-\Lambda^{A'A}}=\frac{1}{c}\,\lambda^{AA'}\,,\qquad \Lambda^{A'B'} = \lambda^{A'B'} \qquad \Theta_\mu = \theta_\mu\,,\label{eq:parresc}
\end{align} we derive the following non-relativistic transformation rules after taking the $c \rightarrow \infty$ limit
\begin{align}
  &\delta \tau_\mu{}^A = \lambda_M \epsilon^A{}_B \tau_\mu{}^B  \,, & \delta e_\mu{}^{A^\prime} &= -\lambda_A{}^{A^\prime} \tau_\mu{}^A + \lambda^{A^\prime}{}_{B^\prime} e_\mu{}^{B^\prime} \,, \nonumber \\
  &\delta b_{\mu\nu} = 2 \partial_{[\mu} \theta_{\nu]} - 2\,\epsilon_{AB}\lambda^A{}_{A^\prime}\,\tau_{[\mu}{}^{B}e_{\nu]}{}^{A^\prime} \,, & \delta \phi &= 0 \,,
\end{align}
where $\lambda_M$, $\lambda^{A^\prime B^\prime}$, $\lambda^{A A^\prime}$ and $\theta_\mu$ are now interpreted as parameters of the longitudinal SO$(1,1)$, transversal SO$(8)$, Galilean boosts and abelian two-form symmetry of the non-relativistic theory. In order to obtain these formulae, it was important that the redefinitions \eqref{eq:R=NR} are invertible. Note that this limit indeed reproduces the correct transformation rule \eqref{eq:deltabgf} of the NR Kalb-Ramond field $b_{\mu\nu}$ under Galilean boosts that was necessary to ensure boost invariance of the string worldsheet action \eqref{curvedindices}. In a similar way, one finds that the projective inverse Vielbeine transform as
\begin{align}
  \delta\tau_A{}^\mu = \lambda_M\,\epsilon_A{}^{B}\tau_B{}^{\mu} + \lambda_A{}^{A^\prime}e_{A^\prime}{}^{\mu}\,, \qquad \qquad \delta e_{A^\prime}{}^\mu = \lambda_{A^\prime}{}^{B^\prime} e_{B^\prime}{}^\mu \,.
\end{align}

The NR limit can similarly be performed on other quantities that are expressed in terms of the fields of relativistic NS-NS gravity. To do this, one plugs the redefinitions \eqref{eq:R=NR} in the quantity of interest, expands the result in powers of $c^{-2}$ and retains only the terms that appear at leading order. In the next subsection, we will apply this procedure to the relativistic NS-NS gravity action. The NR limit of the equations of motion of NS-NS gravity will be considered in section \ref{sec:NRNSNS}.

\subsection{Taking the NR Limit of the NS-NS Gravity Action} \label{ssec:NSNSrel}

The dynamics of the fields of relativistic NS-NS gravity is governed by the following action (in the string frame):
\begin{equation}
  \label{eq:NSNSaction}
  S_{\rm NS-NS} = \frac{1}{2\kappa^2} \int \rmd^{10} x \, E\, \rme^{-2 \Phi} \left(\mathcal{R} + 4\, \partial_\mu \Phi \partial^\mu \Phi - \frac12 \mathcal{H}^2 \right) \,.
\end{equation}
Here, $\kappa$ is the gravitational coupling constant, $E = \mathrm{det}(E_\mu{}^{\hat{A}})$, $\mathcal{R}$ is the Ricci scalar of $G_{\mu\nu}$ and
\begin{equation}
  \label{eq:defHH2}
  \mathcal{H}^2 = \frac{1}{3!} \mathcal{H}_{\mu\nu\rho} \mathcal{H}^{\mu\nu\rho} \,, \qquad \qquad \text{with } \ \mathcal{H}_{\mu\nu\rho} = 3\,\partial_{[\mu} B_{\nu\rho]} \,.
\end{equation}
We refer to Appendix \ref{ssec:LorGeom} for our conventions on Lorentzian geometry.

To take the NR limit of the relativistic NS-NS gravity action \eqref{eq:NSNSaction}, we plug the redefinitions \eqref{eq:R=NR} in \eqref{eq:NSNSaction} and expand the result in powers of $c^{-2}$. The leading order term of the resulting expansion then appears a priori at order $c^2$:
\begin{align} \label{eq:expSNSNS}
    S_{\mathrm{NS-NS}} = c^2\,\accentset{(2)}{S} + c^0\,\accentset{(0)}{S} + c^{-2}\,\accentset{(-2)}{S} + \cdots \,.
\end{align}
Here, the explicit expression for $\accentset{(2)}{S}$ is proportional to
\begin{align}
    \accentset{(2)}{S}\propto \int \rmd^{10}x\,e\,\rme^{-2\phi}\,\left(\accentset{(2)}{\mcR} - \frac14\,\accentset{(2)}{\mathcal{H}}_{AA'B'}\accentset{(2)}{\mathcal{H}}{}^{AA'B'}\right) \,,
\end{align}
where $e=\mathrm{det}(\tau_\mu{}^A, e_\mu{}^{A^\prime})$, $\accentset{(2)}{\mcR}$ is the term at order $c^2$ in the expansion of the Ricci scalar $\mathcal{R}$ and  $\accentset{(2)}{\mathcal{H}}_{\mu\nu\rho}$ is the term at order $c^2$ in the expansion of $\mathcal{H}_{\mu\nu\rho}$.\footnote{{Note that the expression $\accentset{(2)}{\mathcal{H}}_{AA'B'}$ is \emph{not} the term at order $c^2$ in the expansion of $\mathcal{H}_{AA'B'}$. Rather, it denotes the contraction with non-relativistic Vielbeine $\tau_A{}^\mu e_{A^\prime}{}^\nu e_{B^\prime}{}^\rho \accentset{(2)}{\mathcal{H}}_{\mu\nu\rho}$. See appendices \ref{ssec:indexconventions} and \ref{sec:SGGinlimit} for more details.}} Both terms in $\accentset{(2)}{S}$ are separately non-zero. Using the explicit expressions (see also Appendix \ref{sec:SGGinlimit})
\begin{equation}
  \accentset{(2)}{\mcR} = -\eta_{AB}\tau_{A'B'}{}^A\tau^{A'B'B} \,, \ \ \ \accentset{(2)}{\mathcal{H}}_{AA'B'} = 2\,\epsilon_{AB}\tau_{A'B'}{}^B \ \ \ (\text{with } \tau_{A^\prime B^\prime}{}^A = e_{A^\prime}{}^\mu e_{B^\prime}{}^\nu \partial_{[\mu} \tau_{\nu]}{}^A) \,,
\end{equation}
one however sees that the contributions of the two terms in $\accentset{(2)}{S}$ exactly cancel. This is a non-trivial cancellation between the Ricci scalar and the kinetic term of the KR field, that mirrors the cancellation, mentioned under \eqref{eq:redefsm}, in the string worldsheet action.

The upshot of this cancellation is that the actual leading order term in the expansion \eqref{eq:expSNSNS} is $\accentset{(0)}{S}$, appearing at order $c^0$. This term can be written in terms of geometric quantities, that characterize a non-Lorentzian geometry that we call {`torsional string Newton Cartan geometry' (TSNC)}. In TSNC geometry, we define spin connections $\omega_\mu$, $\omega_\mu{}^{A^\prime B^\prime}$, $\omega_\mu{}^{A A^\prime}$ for SO$(1,1)$, SO$(8)$ and Galilean boosts as well as a field $b_\mu$ that we will call `the dilatation connection'. These connections are defined in terms of $\tau_\mu{}^A$, $e_\mu{}^{A^\prime}$, $b_{\mu\nu}$ and $\phi$ as follows:
\begin{subequations}\label{eq:allconnections}
    \begin{align}
    b_\mu &=
            e_\mu{}^{A'} e_{A^\prime}{}^\nu \tau_A{}^\rho  \partial_{[\nu} \tau_{\rho]}{}^A +  \tau_\mu{}^A  \tau_A{}^\nu  \partial_\nu \phi\,,\label{eq:dilsc}\\
    \omega_\mu &=
            \left( \tau^{A \nu}  \partial_{[\mu} \tau_{\nu]}{}^{B} -  \frac12 \tau_\mu{}^C \tau^{A \nu}  \tau^{B \rho} \partial_{[\nu} \tau_{\rho] C} \right) \epsilon_{AB} -  \epsilon_{AB}  \tau_\mu{}^A   \tau^{B \nu}  \partial_\nu \phi\,,\label{eq:longsc}\\
    \omega_\mu{}^{AA'} &=
            - \tau^{A \nu} \partial_{[\mu} e_{\nu]}{}^{A'} + e_{\mu B'}  \tau^{A \nu}  e^{A^\prime \rho} \partial_{[\nu} e_{\rho]}{}^{B'} + \frac32 \epsilon^{AB} \tau_B{}^{\nu}  e^{A^\prime \rho} \partial_{[\mu} b_{\nu\rho]} + \tau_{\mu B} W^{A B A'}\,,\label{eq:galboostsc}\\
    \omega_\mu{}^{A'B'} &=
            -2 e^{[A^\prime|\nu|} \partial_{[\mu} e_{\nu]}{}^{B']} +  e_{\mu C'}  e^{A^\prime \nu} e^{B^\prime \rho} \partial_{[\nu} e_{\rho]}{}^{C'}  -  \frac32 \tau_\mu{}^A \epsilon_{AB} \tau^{B \nu}  e^{A^\prime \rho}  e^{B^\prime \sigma} \partial_{[\nu} b_{\rho\sigma]} \,,\label{transsc}
    \end{align}
\end{subequations}
Here, $W^{AB A^\prime}$ is a tensor that is symmetric traceless in the $AB$ indices, but is otherwise arbitrary.\footnote{As explained in Appendix \ref{sec:stringGalileigeometry}, the connections \eqref{eq:allconnections} are found as solutions of conventional constraints. The presence of the arbitrary tensor $W^{A B A^\prime}$ then indicates that the imposed conventional constraints do not suffice to determine all boost connection components uniquely. It should be noted that the presence of this term is irrelevant for what follows, as it drops out of the NR action and equations of motion.} We refer to Appendix \ref{sec:stringGalileigeometry} for more information on the definitions of these and related TSNC geometric quantities. Using these connections, the leading order term $\accentset{(0)}{S}$ can then be written as
\begin{align} \label{eq:NRaction1}
 \accentset{(0)}{S} = \frac{1}{2\,\kappa^2}\int \rmd^{10}x \,e\,\rme^{-2\,\phi}\bigg(&\rmR(J)+4\,\partial_{A'}\phi\,\partial^{A'}\phi-\frac{1}{12}\,h_{A'B'C'}h^{A'B'C'} \notag\\
 &-4\,\mathcal D _{A'}b^{A'} - 4\,b_{A'}b^{A'}- 4\,\tau_{A'\{AB\}}\tau^{A'\{AB\}}  \bigg)\,.
\end{align}Here, $\tau_{\mu\nu}{}^A = \partial_{[\mu} \tau_{\nu]}{}^A$, $h_{\mu\nu\rho} = 3 \partial_{[\mu} b_{\nu \rho]}$ and we have turned curved indices into flat ones using $\tau_\mu{}^A$, $e_\mu{}^{A^\prime}$, $\tau_A{}^\mu$, $e_{A^\prime}{}^\mu$, as detailed in Appendix \ref{ssec:indexconversions}. The curvature scalar $\rmR(J)$ and derivative $\mathcal D_\mu b^{A^\prime}$ are explicitly given by
\begin{subequations}
    \begin{align}
        \rmR(J) &= -2\,e_{A^\prime}{}^\mu e_{B^\prime}{}^\nu \left(\partial_{[\mu} \omega_{\nu]}{}^{A'B'} + \omega_{[\mu}{}^{A'C'}\omega_{\nu]}{}^{B'}{}_{C'}\right) {- 4\,\omega^{A'BB'}\tau_{A'B'B}} \,,  \\ \label{eq:derbexpl}
        \mathcal D_\mu b^{A'} &= \partial_\mu b^{A'} - \omega_\mu{}^{A'B'}b_{B'} - \omega_\mu{}^{AB'}\tau_{A'B'A}\,.
    \end{align}
\end{subequations}
Note that $\mathcal D_\mu$ is covariant with respect to $\mathrm{SO}(1,1)\times \mathrm{SO}(8)$ and Galilean boosts. We refer to Appendix \ref{sec:SGGinlimit} for details on how \eqref{eq:NRaction1} is obtained.

Like the Stueckelberg gauge-fixed string worldsheet action \eqref{curvedindices}, this action \eqref{eq:NRaction1} is invariant under String Galilei symmetries and dilatations that act as
\begin{align}
  &\delta\tau_\mu{}^{A} = \lambda_M\,\epsilon^A{}_B\tau_\mu{}^B + \lambda_D \tau_\mu{}^A \,, & \delta e_\mu{}^{A'} &= \lambda^{A'}{}_{B'}e_\mu{}^{B'} - \lambda_A{}^{A'}\tau_\mu{}^{A}\,, \nonumber \\
  &\delta b_{\mu\nu} = - 2\,\epsilon_{AB}\lambda^A{}_{A^\prime}\,\tau_{[\mu}{}^{B}e_{\nu]}{}^{A^\prime} \,, &\delta \phi &= \lambda_D \,,
\end{align}
where $\lambda_M$, $\lambda^{A^\prime B^\prime}$, $\lambda^{A A^\prime}$, $\lambda_D$ are the parameters of SO$(1,1)$, SO$(8)$, Galilean boosts and dilatations resp. Note that the fields $\omega_\mu$, $\omega_\mu{}^{A^\prime B^\prime}$ and $\omega_\mu{}^{A A^\prime}$, defined in \eqref{eq:allconnections}, then indeed transform as connections for SO$(1,1)$, SO$(8)$ and Galilean boosts resp., while $b_\mu$ transforms as a gauge field under dilatations, as anticipated by calling it `dilatation connection'.

The invariance under String Galilei symmetries is not surprising, since \eqref{eq:NRaction1} appears as the leading order term in an expansion in powers of $c^{-2}$. As such, it is guaranteed to be invariant under the NR limit of SO$(1,9)$ local Lorentz symmetries, i.e., under the String Galilei symmetries. The invariance under dilatations is more surprising. Like in the NR limit of the string worldsheet action, it appears here as an emergent symmetry.

One can however rewrite \eqref{eq:NRaction1} in a way that is manifestly dilatation invariant. This rewriting is achieved by partially integrating the $e\,\rme^{-2\, \phi} \mathcal D_{A^\prime} b^{A^\prime}$ term, using \eqref{eq:partorpe}. Doing this, one finds that \eqref{eq:NRaction1} is {equivalent to}
\begin{align}\label{eq:NRaction2}
    S_{\mathrm{NR}}[\tau_\mu{}^A,e_\mu{}^{A^\prime},b_{\mu\nu},\phi] = \frac{1}{2\,\kappa^2}\int \rmd^{10}x\,e\,\rme^{-2\,\phi}\bigg(&\rmR(J) +4\, \nabla_{A^\prime} \phi\nabla^{A^\prime} \phi-\frac{1}{12}\,h_{A'B'C'}h^{A'B'C'}\notag\\
    &- 4\,\tau_{A'\{AB\}}\tau^{A'\{AB\}} {+4\,\omega^{A'BB'}\tau_{A'B'B}}  \bigg) \,,
\end{align}{up to a boundary term $-4\,\partial_\mu\big(e\,\rme^{-2\phi}e_{A'}{}^\mu\,b^{A'}\big)$.}
Manifest dilatation invariance is then achieved by virtue of the fact that the dependent field $b_\mu$ corresponds to a gauge field for dilatations (see \eqref{eq:deltaomega1,18}) and that $\nabla_\mu\phi = \partial_\mu\phi - b_\mu$ is thus dilatation invariant. {The occurrence of an explicit boost spin connection in eq. \eqref{eq:NRaction2} is indicative of the fact that the Lagrangian is boost invariant only up to a total derivative. This is also clear from the form of the boundary term. To summarize, we present two physically equivalent ways of writing the NR action---one, \eqref{eq:NRaction1}, in which boost symmetry is manifest and dilatation symmetry is not, and one, \eqref{eq:NRaction2}, where dilatation symmetry is manifest but boost symmetry is not. In order to distinguish the two, we will continue to use $\accentset{(0)}{S}$ for \eqref{eq:NRaction1} and $S_{NR}$ for \eqref{eq:NRaction2}, even though they give rise to the same equations of motion.}

The non-relativistic action \eqref{eq:NRaction2} contains the background fields of non-relativistic string theory. This should be contrasted to the situation that would occur, had the two contributions in $\accentset{(2)}{S}$ not cancelled. In that case, one would have ended up with a non-relativistic action (namely $\accentset{(2)}{S}$) that only contains $\tau_\mu{}^A$.

Let us now look at the equations of motion that are derived from \eqref{eq:NRaction2}. We denote the equations of motion of $\tau_\mu{}^A$, $e_\mu{}^{A^\prime}$, $b_{\mu\nu}$ and $\phi$ by $\langle\tau \rangle_A{}^\mu$, $\langle e\rangle_{A^\prime}{}^\mu$, $\langle b \rangle^{\mu\nu}$ and $\langle\phi\rangle$ and define them via the following variation:
\begin{align}
  \delta\,S_{\mathrm{NR}}= \frac{1}{2\,\kappa^2}\int \rmd^{10}x\,e\,\rme^{-2\phi}\bigg(&\langle\tau\rangle_A{}^\mu\delta\tau_\mu{}^A + \langle e \rangle_{A'}{}^\mu\delta e_\mu{}^{A'}
    -8\,\langle\phi\rangle\delta\phi  + \frac12\,\langle b \rangle^{\mu\nu}\delta b_{\mu\nu}\bigg)\,.
\end{align}
Here, the pre-factors have been chosen for later convenience. In total, the equations of motion $\langle\tau \rangle_A{}^\mu$, $\langle e\rangle_{A^\prime}{}^\mu$, $\langle b \rangle^{\mu\nu}$ and $\langle\phi\rangle$ consist of 20 + 80 + 45 + 1 = 146 components. Not all of these components are independent however. Indeed, the invariance of the action \eqref{eq:NRaction2} under String Galilei symmetries and dilatations implies the following algebraic relations (Noether identities):
\begin{align} \label{eq:NRNoetherids}
    & \langle \tau \rangle_{[AB]}=0\,,\quad \ \ \ \langle e \rangle_{[A'B']}=0\,,\quad \ \ \ \langle e \rangle_{A'}{}^\mu\tau_{\mu A}  +\epsilon_A{}^B\, \langle b\rangle_{BA'} =0\,,\quad \ \ \
    \langle \tau \rangle_A{}^\mu\tau_\mu{}^A = 8\,\langle\phi\rangle\,.
\end{align}
The first three relations are the Noether identities for the SO$(1,1)$, SO$(8)$ and Galilean boost symmetries, while the last one corresponds to the Noether identity of dilatations. In total, we have $1 + 28 + 16 + 1 = 46$ such algebraic relations. Subtracting this number from the total number of 146 of equations, contained in $\langle\tau\rangle_A{}^\mu$, $\langle e \rangle_{A^\prime}{}^\mu$, $\langle b\rangle^{\mu\nu}$, $\langle\phi\rangle$, we are left with 100 equations.

In order to give the equations of motion $\langle\tau\rangle_A{}^\mu$, $\langle e \rangle_{A^\prime}{}^\mu$, $\langle b\rangle^{\mu\nu}$, $\langle\phi\rangle$ explicitly, we turn the curved indices into flat ($A$ or $A^\prime$) ones, and decompose the resulting tensors into irreducible representations of $\mathrm{SO}(1,1)\times\mathrm{SO}(8)$. {Strictly speaking, this is not necessary at this point but turns out to be convenient when comparing with the equations of motion in the next section, and the beta functions in section \ref{sec:betafunc}. We rename the representations as follows:}
\begin{align} \label{eq:nractioneomsirreps}
    &\langle S_-\rangle \epsilon^{AB} \equiv 2\,\langle b\rangle ^{AB}\,, &&\langle G\rangle _{\{AB\}} \equiv -\frac12 \langle \tau\rangle _{\{AB\}}\,, \nonumber \\
    &\langle V_-\rangle _{AA'} \equiv -\frac12 \langle \tau\rangle _{AA'}\,,    &&\langle G\rangle _{A'B'} \equiv 2\,\delta_{A'B'}\,\langle\phi\rangle- \frac12\langle e\rangle _{(A'B')}\,, \nonumber \\
    &\langle V_+\rangle _{AA'} \equiv -\langle e\rangle _{A'A} = \epsilon_A{}^B\langle b\rangle _{BA'}\,, &&\langle \Phi \rangle \equiv \langle \phi\rangle  = \frac18 \eta^{AB}\langle \tau\rangle _{AB}\,, \nonumber \\
  & \langle B \rangle_{A^\prime B^\prime} \equiv \langle b \rangle_{A^\prime B^\prime} \,.
\end{align}
In terms of these irreducible representations, $\langle\tau\rangle_A{}^\mu$, $\langle e \rangle_{A^\prime}{}^\mu$ and $\langle b\rangle^{\mu\nu}$ are decomposed as follows:
\begin{subequations} \label{eq:nractioneomsdecomp}
    \begin{align}
        \langle \tau\rangle _A{}^\mu &
                                = 4\,\tau_A{}^\mu\,\langle \Phi\rangle  -2\,\tau^{B \mu}\,\langle G\rangle _{\{AB\}} -2\, e^{A^\prime \mu}\,\langle V_-\rangle _{AA'}\,,\\
        \langle e\rangle _{A'}{}^\mu &
                                = 4\,e_{A'}{}^\mu\,\langle \Phi\rangle-2\,e^{B' \mu}\langle G\rangle _{A'B'} - \tau^{A \mu}\langle V_+\rangle _{AA'}\,,\\
        \langle b\rangle ^{\mu\nu} &
                                = \frac12\,\epsilon^{AB}\tau_A{}^\mu\tau_B{}^\nu\,\langle S_-\rangle  + e_{A'}{}^\mu e_{B'}{}^\nu\,\langle B\rangle ^{A'B'} + 2\,\tau_A{}^{[ \mu} e^{|A^\prime|\nu]}\epsilon^{AB}\,\langle V_+\rangle _{BA'} \,.
\end{align}
\end{subequations}
Note that we have taken the redundancy due to the Noether identities \eqref{eq:NRNoetherids} into account in \eqref{eq:nractioneomsirreps} and \eqref{eq:nractioneomsdecomp}. The 100 independent equations of motion, derived from \eqref{eq:NRaction2} are thus given by
\begin{align} \label{eq:allNRactioneoms}
  \langle S_- \rangle &= 0 \,, &\langle V_\pm \rangle_{A A^\prime} &= 0 \,, & \langle G \rangle_{\{AB\}} &= 0 \,,  \nonumber \\
  \langle \Phi \rangle &= 0 \,,& \langle G \rangle_{A^\prime B^\prime} &= 0 \,, & \langle B \rangle_{A^\prime B^\prime} &= 0 \,.
\end{align}
These are then explicitly found to be
\begin{subequations}\label{eq:nrEOM}
    \begin{align}
        \langle \Phi\rangle   &
                                =\nabla^{A'}\nabla_{A'}\phi - \big(\nabla_{A'}\phi \big)^2 + \frac14\,\rmR(J) - \frac{1}{48}\,h_{A'B'C'}h^{A'B'C'} - \tau_{A'\{AB\}}\tau^{A'\{AB\}}\,,\label{eq:phi}\\
        \langle G\rangle _{\{AB\}} &
                                = -2\,\big(\nabla_{B'} - 2\,(\nabla_{B'}\phi)\big)\tau^{B'}{}_{\{AB\}}\,,\label{eq:STL}\\
        \langle V_-\rangle _{AA'} &
                                = \rmR_{C'A}(J)_{A'}{}^{C'} + 2\,\nabla_A\nabla_{A'}\phi + 2\,\nabla^B\tau_{A'\{AB\}}\,,\label{eq:V-}\\
        \langle G\rangle _{A'B'} &
                                = \rmR_{C'(A'}(J)_{B')}{}^{C'} +2\,\nabla_{(A'}\nabla_{B')}\phi -  \frac14\,h_{A'C'D'}h_{B'}{}^{C'D'} - 4\,\tau_{A'\{AB\}}\tau_{B'}{}^{\{AB\}}\,,\label{eq:GA'B'}\\
        \langle V_+\rangle _{AA'} &
                                = -2\,\big(\nabla_{B'} -2\,( \nabla_{B'}\phi)\big)\,\tau^{B'}{}_{A'A} - 4\,\tau^{B'}{}_{\{AB\}}\tau_{B'A'}{}^B - \epsilon_{AB}\,h_{A'B'C'}\tau^{B'C'B}\,,\label{eq:V+}\\
        \langle S_-\rangle  &
                                = 4\,\tau_{A'B'C}\tau^{A'B'C}\,,\label{eq:S-}\\
        \langle B\rangle _{A'B'} &
                                = \big(\nabla_{C'} - 2\,(\nabla_{C'}\phi)\big) h^{C'}{}_{A'B'} + 2\,\epsilon^{AB}\,\nabla_A\tau_{A'B'B}\,,\label{eq:BA'B'}
    \end{align}
\end{subequations}
where the derivative $\nabla_\mu$ is covariantized with respect to dilatations, $\mathrm{SO}(1,1)\times \mathrm{SO}(8)$ and Galilean boosts. The curvature $R_{\mu\nu}(J)^{A^\prime B^\prime}$ is defined in eqs. \eqref{eq:curv2forms}. {One can show that none of these equations depends on the undetermined symmetric traceless part of the boost spin connection \eqref{eq:galboostsc} $W_\mu{}^{AA'}$.}

Note that none of the above equations \eqref{eq:nrEOM} contains a term of the form $\partial_{A^\prime} \partial^{A^\prime} b_{01}$ at the linearized level. Since $b_{01}$ can be identified with the Newton potential (in the Stueckelberg gauge fixed formalism with $m_\mu{}^A = 0$ that we are working in), this means that none of the equations \eqref{eq:nrEOM} can be interpreted as a covariant version of a Poisson-type equation of NR gravity. In the next section, we will consider how to take the NR limit of the equations of motion of relativistic NS-NS gravity directly. As we will see, this limit can be defined such that it not only reproduces the equations \eqref{eq:nrEOM}, but also leads to an extra Poisson-type equation.

\section{The NR Limit of the NS-NS Gravity Equations of Motion} \label{sec:NRNSNS}

In this section, we will discuss the results of applying the NR limit \eqref{eq:R=NR} to the equations of motion of relativistic NS-NS gravity. We will first discuss how these equations of motion can be reorganized, such that their NR limit can be taken in an appropriate manner. Next, we will discuss the NR equations that result from the limit and compare them to the equations of motion \eqref{eq:nrEOM} that are derived from the NR action \eqref{eq:NRaction2}.

We will denote the equations of motion for the fields $G_{\mu\nu}$, $B_{\mu\nu}$ and $\Phi$ of relativistic NS-NS gravity by $[G]_{\mu\nu}$, $[B]_{\mu\nu}$ and $[\Phi]$ resp. They are derived from the action \eqref{eq:NSNSaction} and are given by\footnote{To be precise, the equations of motion, derived from the action \eqref{eq:NSNSaction}, are equivalent to the equations, given in \eqref{eq:EOM}. The action \eqref{eq:NSNSaction} leads to $[B]_{\mu\nu}=0$ and $[\Phi]=0$ as equations of motion for $B_{\mu\nu}$ and $\Phi$, while it gives $[G]_{\mu\nu} - 2 G_{\mu\nu} [\Phi]=0$ as equation of motion for $G_{\mu\nu}$.}
\begin{subequations}\label{eq:EOM}
    \begin{align}
    [G]_{\mu\nu} & \equiv \mathcal R_{\mu\nu} + 2\,\nabla_\mu\partial_\nu\Phi - \frac14 \mathcal{H}_{\mu\rho\sigma} \mathcal{H}_\nu{}^{\rho\sigma} = 0 \,, \label{eq:EOM1}\\
    [B]_{\mu\nu} & \equiv \nabla^\rho \mathcal{H}_{\rho\mu\nu} - 2\lr\partial^\rho\Phi\rr \mathcal{H}_{\rho\mu\nu} = 0 \,, \label{eq:EOM2} \\
    [\Phi] &\equiv \nabla^\mu\partial_\mu\Phi + \frac14 \mathcal{R}- \partial^\mu\Phi\partial_\mu\Phi - \frac{1}{48}\, \mathcal{H}_{\mu\nu\rho} \mathcal{H}^{\mu\nu\rho} = 0 \,. \label{eq:EOM3}
\end{align}
\end{subequations}
Note that these constitute 55 + 45 + 1 = 101 relativistic equations of motion.

In string theory, these equations \eqref{eq:EOM} also ensure that scale invariance of the string worldsheet action is maintained at the quantum mechanical level. Indeed, in the Polyakov action for the relativistic string, $G_{\mu\nu}$, $B_{\mu\nu}$ and $\Phi$ can be viewed as coupling constants. Quantum scale invariance of the string action, then requires that the beta functions $\beta^G_{\mu\nu}$, $\beta^B_{\mu\nu}$ and $\beta^\Phi$ of $G_{\mu\nu}$, $B_{\mu\nu}$ and $\Phi$ vanish.

We can then take the NR limit of the equations of motion \eqref{eq:EOM}, by plugging \eqref{eq:R=NR} in \eqref{eq:EOM}, expanding the resulting equations in powers of $c^{-2}$ and retaining only the terms at leading order as NR equations of motion. If one does this for the 101 equations of motion, as they are given in \eqref{eq:EOM}, one finds that some of them give rise to the same NR equation. In particular, one finds that the leading order terms in the expansion of {$[G]_{AA'}$ and $\epsilon_A{}^B[B]_{BA'}$ are proportional to each other.\footnote{Here, we have turned curved indices on components of \eqref{eq:EOM} into flat indices using the relativistic (inverse) Vielbeine $E_A{}^\mu$, $E_{A^\prime}{}^\mu$. For example $[G]_{AA'} = E_A{}^\mu E_{A^\prime}{}^\nu [G]_{\mu\nu}$. For more details, see Appendix \ref{ssec:indexconversions}.} Remarkably, for $\eta^{AB}[G]_{AB}$ and $\epsilon^{AB}[B]_{AB}$, both the terms at leading \emph{and} at subleading order are proportional to each other.}
{To avoid this redundancy, one can instead consider the following $101$ equations}
\begin{align} \label{eq:2,8EOM}
  & [S_\pm] \equiv \eta^{AB}[ G]_{AB} \pm\frac12\,\epsilon^{AB}[ B]_{AB} = 0 \,, \qquad \qquad [V_\pm]_{AA'} \equiv  [G]_{AA'}\pm\frac12\,\epsilon_A{}^B\,[B]_{BA'} = 0 \,, \nonumber \\
  & [G]_{\{A B\}} = 0 \,, \qquad \qquad [G]_{A^\prime B^\prime} = 0 \,, \qquad \qquad  [\Phi] = 0 \,, \qquad \qquad [B]_{A^\prime B^\prime} = 0 \,.
\end{align}
Plugging in the redefinitions \eqref{eq:R=NR} into \eqref{eq:2,8EOM} and expanding the resulting equations in powers of $c^{-2}$, we then find that leading order terms for different equations occur at different powers of $c$. {We will use the following notation
\begin{align}\label{leadingordernotation}
    [X] = c^n\langle X\rangle + \mathcal{O}(c^{n-2})\,,
\end{align}
to denote the expansion of the relativistic equations $[X]$ as in \eqref{eq:2,8EOM}, and the terms at leading order $n$ as $\langle X\rangle$. For example, one can show that $[S_-]$ has $n=+2$, whereas $[S_+]$ has $n=- 2$. The fact that the singlet equations $[S_\pm]$ have leading orders separated by a factor $c^4$ is remarkable and has important consequences for the structure of the non-relativistic theory.\footnote{{Similar observations have been made in the context of non-relativistic expansions of General Relativity \cite{Hansen:2019vqf}.}}}

For future reference, we indicate here how all components of the above equations of motion transform into each other under the boosts (with parameters $\Lambda^{A A^\prime}$) of SO$(1,9)$:
\begin{align}\label{eq:boost}
    &\delta [\Phi] = 0\,, \qquad\quad
    \delta [S_\pm] = 2\, \Lambda^{AA'}[V_\mp]_{AA'}\,,\qquad \quad~
    \delta [G]_{\{AB\}} = \Lambda_{\{A}{}^{A'}[V_+]_{B\}A'} + \Lambda_{\{A}{}^{A'}[V_-]_{B\}A'}\,,\notag\\
    &\delta[V_+]_{AA'} = \Lambda_A{}^{B'}[G]_{A'B'} -\frac12\,\epsilon_A{}^B\Lambda_B{}^{B'}[B]_{A'B'} - \Lambda^B{}_{A'}[G]_{\{AB\}} - \frac12\,\Lambda_{AA'}[S_-]\,,\notag\\
    &\delta[V_-]_{AA'} = \Lambda_A{}^{B'}[G]_{A'B'} +\frac12\,\epsilon_A{}^B\Lambda_B{}^{B'}[B]_{A'B'} - \Lambda^B{}_{A'}[G]_{\{AB\}} - \frac12\,\Lambda_{AA'}[S_+]\,,\\
    &\delta [G]_{A'B'} = -\Lambda^A{}_{(A'}\big([V_+]_{|A|B')}+[V_-]_{|A|B')}\big)\,,\qquad \delta [B]_{A'B'} = - 2\,\Lambda^A{}_{[A'}\epsilon_{|A}{}^B\big([V_+]_{B|B']}-[V_-]_{B|B']}\big)\,.\notag
\end{align}

The NR limit is then obtained by retaining only {the leading order terms $\langle X\rangle$ in eqs. \eqref{eq:2,8EOM}}. In this way, we obtain 101 NR equations as the NR limit of the relativistic NS-NS gravity equations of motion. These 101 NR equations are given by the 100 equations given in eqs. \eqref{eq:allNRactioneoms}, \eqref{eq:nrEOM}, as well as the extra equation:
\begin{align} \label{eq:extraeq}
    \langle S_+\rangle \equiv -\tau_A{}^\mu e_{A^\prime}{}^\nu \rmR_{\mu\nu}(G)^{AA'} - \epsilon^{AB}\tau_A{}^\mu \tau_B{}^\nu \rmR_{\mu\nu}(M) = 0\,,
\end{align}
using the notation introduced in eq. \eqref{leadingordernotation}. $\rmR_{\mu\nu}(G)^{A A^\prime}$ and $\rmR_{\mu\nu}(M)$ are defined in \eqref{eq:curv2forms}.

{We have seen that the non-relativistic action \eqref{eq:NRaction2} is invariant under dilatations. Consequently, the equations of motion also transform covariantly under transformations \eqref{localD}. It turns out that the leading order $n$ at which the equations occur in the expansion \eqref{leadingordernotation} is the dilatation weight of the corresponding non-relativistic equation of motion, i.e., $\delta_D\langle X\rangle = n\,\lambda_D\langle X\rangle$. We summarize all the dilatation weights in table \ref{fig:ordEOM}.
\begin{table}[h]
    \centering
    \begin{tabular}{c|cccccccc}
         $\langle X\rangle$ & $\langle\Phi\rangle$ & $\langle S_-\rangle$ & $\langle S_+\rangle$ & $\langle G \rangle_{\{AB\}}$ & $\langle V_-\rangle_{AA'}$ & $\langle V_+\rangle_{AA'}$ & $\langle G \rangle_{A'B'}$ & $\langle B\rangle_{A'B'}$ \\\hline
         $n$ & $0$ & $2$ & $-2$ & $0$ & $-1$ & $+1$ & $0$ & $0$
         \end{tabular}
    \caption{Dilatation weights of the equations of motion $\delta_D\langle X\rangle = n\,\lambda_D\langle X\rangle$}
    \label{fig:ordEOM}
\end{table}}

Note that the NR action \eqref{eq:NRaction2} gives rise to one equation of motion less than the relativistic NS-NS gravity action \eqref{eq:NSNSaction}. This discrepancy is consistent with the fact that the NR action \eqref{eq:NRaction2} enjoys this extra dilatation symmetry, that emerges after taking the limit. 
The additional equation \eqref{eq:extraeq} that is not obtained from the NR action \eqref{eq:NRaction2} corresponds to (a covariant version of) the Poisson equation of the NR gravity theory that is described by NR string theory. Indeed, the expression $\langle S_+ \rangle$ contains a term $\partial_{A'}\partial^{A'}\,b_{01}$, where $b_{01}$ corresponds to the Newton potential, in the formulation with $m_\mu{}^A = 0$ that we are currently using.

{The emergence of a dilatation symmetry explains the discrepancy in the number of equations of motion. However, it does not explain which equation is lost when restricting to the limit at the level of the NS-NS action. The relevant equation has to be an $\mathrm{SO}(1,1)\times \mathrm{SO}(8)$ singlet---leaving eqs. \eqref{eq:S-}, \eqref{eq:phi}, and \eqref{eq:extraeq} as options. Let us now see why it is the Poisson equation $\langle S_+\rangle$ that does not---and cannot---follow from \eqref{eq:NRaction2}. Recall that the action has dilatation weight zero. Hence every equation of motion corresponds to a field component of opposite dilatation weight. The non-linear equation $\langle S_-\rangle$, for example, has $n=+2$ and follows as the field equation of $\epsilon^{AB} b_{AB}$ which has weight $n=-2$. Using this argument and the fact that the Poisson equation has $n=-2$, we see that it should correspond to an $\mathrm{SO}(1,1) \times \mathrm{SO}(8)$ singlet field of dilatation weight $n=+2$. However, no such field component exists in our theory. Hence, with the field content at hand, it is impossible to derive $\langle S_+\rangle$ from a variational principle compatible with dilatations.}

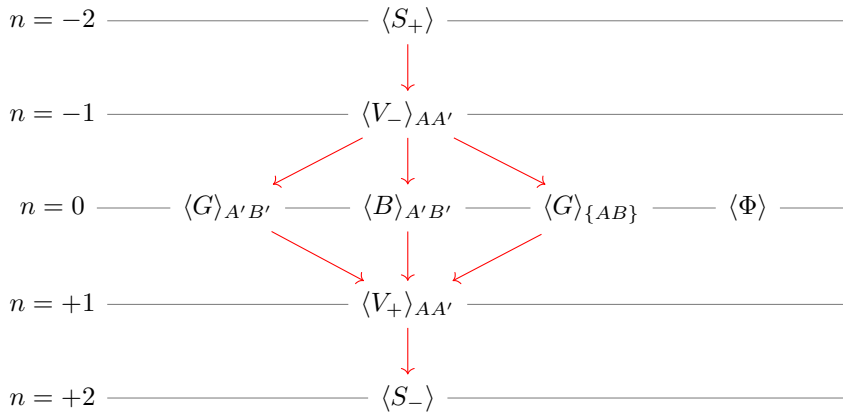
\begin{figure}[h]
\centering
\begin{tikzcd}
  n=-2 \arrow[rr,dash, gray] & & \langle S_+\rangle \arrow[d, red] \arrow[rrr, dash, gray] & & & {}\\
  n=-1\arrow[rr,dash,gray, start anchor=east] & & \langle V_-\rangle_{AA'}\arrow[dl, red]\arrow[d,red]\arrow[dr,red]\arrow[rrr,dash,gray] & & & {}\\
  n=0 &\arrow[l,dash, gray] \langle G\rangle_{A'B'}\arrow[dr, red]\arrow[r,dash, gray] & \langle B\rangle_{A'B'}\arrow[d,red]\arrow[r,dash, gray] & \langle G\rangle_{\{AB\}}\arrow[dl,red]\arrow[r,dash, gray] &\langle\Phi\rangle \arrow[r,dash, gray] & {}\\
  n=+1\arrow[rr,dash,gray, start anchor=east] & & \langle V_+\rangle_{AA'}\arrow[d,red]\arrow[rrr,dash,gray] & & & {}\\
  n=+2\arrow[rr,dash, gray, start anchor=east] & & \langle S_-\rangle \arrow[rrr, dash, gray] & & & {}
\end{tikzcd}
\caption{This diagram summarizes how the equations of motion transform under String Galilei and dilatation symmetries. The index structure indicates the $\mathrm{SO}(1,1)\times\mathrm{SO}(8)$ representations and the layers denote the dilatation weights $\delta_D\langle X\rangle = n\,\lambda_D\langle X\rangle$. Under Galilean boosts, the equations of motion form a reducible, but indecomposable representation. This is indicated by the red arrows.}
\label{tab:rep-th}
\end{figure}

Since the 100 NR equations of motion \eqref{eq:allNRactioneoms} can be obtained from varying a String Galilei invariant action, one finds that they form a representation of the String Galilei symmetries. They transform under $\mathrm{SO}(1,1) \times \mathrm{SO}(8)$ as indicated by their index structure and their transformation rules under Galilean boosts can be inferred from \eqref{eq:boost}:
\begin{align}\label{eq:EOMrep}
    &\delta\langle\Phi\rangle=0\,, \hskip 4.2truecm \delta\langle S_-\rangle = 0\,,\notag\\
    &\delta\langle G \rangle_{\{AB\}} = \lambda_{\{A}{}^{A'}\langle V_+ \rangle_{B\}A'}\,,\hskip 1.3truecm\delta\langle V_+ \rangle_{AA'} = -\frac12\,\lambda_{AA'} \langle S_- \rangle\,,\\
    &\delta\langle V_-\rangle_{AA'} = \lambda_A{}^{B'} \langle G \rangle _{A'B'} +\frac12\,\epsilon_A{}^B \lambda_B{}^{B'} \langle b \rangle_{A'B'} - \lambda^B{}_{A'} \langle G\rangle_{\{AB\}}\,,\notag\\
    &\delta\langle G \rangle_{A'B'} = -\lambda^A{}_{(A'} \langle V_+ \rangle_{|A|B')}\,,\hskip 1.1truecm \delta\langle b\rangle_{A'B'} = -2\,\lambda^A{}_{[A'} \langle V_+\rangle_{|A|B']}\,.\notag
\end{align}
Since, $\langle S_+ \rangle$ transforms under Galilean boosts and dilatations as
\begin{align} \label{eq:deltaSpnr}
    \delta\,\langle S_+\rangle = -2\,\lambda_D\,\langle S_+\rangle + 2\,\lambda^{AA'}\langle V_-\rangle_{AA'}\,,
\end{align}
we see that adding \eqref{eq:extraeq} to the set of 100 equations of motion obtained from the action \eqref{eq:NRaction2}, gives a consistent set of 101 equations of motion that transform as a representation of String Galilei and dilatation symmetries, according to \eqref{eq:EOMrep} and \eqref{eq:deltaSpnr}. {The dilaton equation is a singlet under all the symmetries. The remaining set of $100$ equations forms a reducible, but indecomposable representation. Reducible means that the equations of motion contain smaller sets of equations that are closed under the symmetries of the theory. Indecomposable, on the other hand, means that the subrepresentations cannot be truncated consistently. For example, the nonlinear singlet equation $\langle S_-\rangle$ is inert under Galilean boosts---yet cannot be omitted since $\delta\langle V_+\rangle = -1/2\,\lambda_{AA'}\langle S_-\rangle$. This also demonstrates the special status of the Poisson equation \eqref{eq:extraeq} as it requires all the other equations \eqref{eq:nrEOM} to form a closed set under Galilei boosts. In other words, one could start from the Poisson equation $\langle S_+\rangle$, vary it under Galilean boost, and thereby generate the full set of non-relativistic equations.  The general structure of the reducibility is in correspondence with the dilatation weights. This allows for a schematic summary of the representation theory of the equations of motion, given in figure \ref{tab:rep-th}.}

\section{Comparison with the Beta Functions of NR String Theory}\label{sec:betafunc}

Let us finally comment on the relation between the NR limit of NS-NS gravity, discussed above, and the beta functions of NR bosonic string theory. The latter were calculated in \cite{Gomis:2019zyu,Yan:2019xsf} for the NR string moving in a SNC geometry with zero torsion, i.e., subject to the constraints:
\begin{align}\label{SNCconstraints}
    \tau_{A'(AB)}=0\,,\qquad\mathrm{and}\qquad\tau_{A'B'}{}^A=0\,.
\end{align}
In \cite{Bergshoeff:2019pij}, it was then shown that these `zero torsion beta functions' are reproduced by a NR limit of the beta functions of the relativistic bosonic string, i.e. of the equations of motion of NS-NS gravity. The limit of \cite{Bergshoeff:2019pij} is similar to the limit discussed in this paper. {However, there is an important difference. The limit of \cite{Bergshoeff:2019pij} was taken in the first-order formalism, in which the relativistic spin connection $\Omega_\mu{}^{\hat{A}\hat{B}}$ is treated as an independent field. 
To perform this first-order limit, the redefinitions \eqref{eq:redefsm} were supplemented with rescalings of the components of $\Omega_\mu{}^{\hat{A}\hat{B}}$ with powers of $c$, not allowing them to be divergent in the limit $c\rightarrow \infty$. The zero torsion constraint \eqref{SNCconstraints} is then reproduced by the NR limit of the conventional constraints.  One could also consider the limit of \cite{Bergshoeff:2019pij} in the second-order formalism. This would lead to the same outcome if one requires the expansion of spin connections to be finite after taking the NR limit.
However, in this paper, we took a NR limit of the NS-NS gravity action and the equations of motion  in the second-order formalism and
arranged things, by canceling divergences when expanding the action or combining equations when expanding the equations of motion, such that the leading order terms are always of the order $c^0$ before taking the NR limit and we could extract the maximum number of NR equations of motion. We showed that one can take the limit by adopting the redefinitions \eqref{eq:R=NR} and expanding the Kalb-Ramond field strength $\mathcal{H}_{\mu\nu\rho}$ and the second-order $\Omega_\mu{}^{\hat{A}\hat{B}}$ accordingly. The leading order terms of $\Omega_\mu{}^{A A^\prime}$ and $\Omega_\mu{}^{A^\prime B^\prime}$ in these expansions appear at one $c^2$-order higher than the order dictated by the rescalings of $\Omega_\mu{}^{A A^\prime}$ and $\Omega_\mu{}^{A^\prime B^\prime}$ in the first-order limit. Naively, this leads to terms that would diverge in the $c \rightarrow \infty$ limit (compared to the first-order limit). However, we noticed that such a divergence is absent, due to a fine-tuned cancellation between the kinetic terms of the metric and of the KR field.}

In contrast to the first-order limit, the second-order NR limit of this paper did not lead to the zero torsion conditions \eqref{SNCconstraints}. Indeed, like the NR string worldsheet actions of section \ref{sec:NRstringaction}, the NR limits of both the action and equations of motion of NS-NS gravity are invariant under the emergent dilatation symmetry \eqref{localD}. The constraints \eqref{SNCconstraints} break this dilatation symmetry and can thus not result from the NR limit considered here. It is worth mentioning that the constraints \eqref{SNCconstraints} can be relaxed to a dilatation invariant set of torsion constraints.
The maximal such set, which defines what we call Dilatation invariant SNC (DSNC) geometry, is given by
\begin{align} \label{TTS}
    \tau_{A'\{AB\}}=0\,,\qquad\mathrm{and}\qquad\tau_{A'B'}{}^A=0\,.
\end{align}
Compared to the SNC constraints \eqref{SNCconstraints}, we have that
\begin{equation}
b_{A'} \equiv \tau_{A'A}{}^A\,,
\end{equation}
which acts like the (transverse components of the) gauge field of local dilatations, is non-zero. 

{Let us now see in how far the equations of motion \eqref{eq:nrEOM}/\eqref{eq:extraeq} obtained here are in agreement with the results from beta function calculations  \cite{Gomis:2019zyu,Yan:2019xsf}. We want to stress that our starting point is different from the one taken in the above references. Here, we have made no a priori assumptions on torsion components---i.e., we work with TSNC geometry. The string sigma model in \cite{Gomis:2019zyu,Yan:2019xsf}, however, is defined on an SNC geometry with the geometric constraints \eqref{SNCconstraints}. Since the starting point is different, we can not compare the results directly. In particular, we find three additional equations that are absent in the beta function analysis
\begin{align}
    \langle S_-\rangle = 0\,,\qquad \langle G\rangle_{\{AB\}}=0\,,\qquad \langle V_+\rangle_{AA'} =0\,.\label{eq:puretorcons}
\end{align}
These impose constraints only on the torsion components and are thus identically satisfied when working with SNC geometry. Here, however, we consider SNC as a solution of eqs. \eqref{eq:puretorcons} rather than an a priori constraint. There are more general solutions of \eqref{eq:puretorcons}, of which DSNC geometry \eqref{TTS} is an example. For the moment, however, let us impose the zero torsion constraints \eqref{SNCconstraints} in order to compare with the beta function calculations. The remaining non-trivial equations
\begin{align}
    \langle S_+\rangle =0\,,\qquad \langle V_-\rangle_{AA'}=0\,,\qquad \langle G\rangle_{A'B'}=0\,,\qquad \langle b\rangle_{A'B'}=0\,,\qquad \langle\Phi\rangle = 0\,,
\end{align}
impose constraints on SNC geometry, the dilaton, and the KR field. We find that these equations are in agreement with the results of \cite{Gomis:2019zyu,Yan:2019xsf}. In other words, the $101$ equations \eqref{eq:nrEOM}/\eqref{eq:extraeq} encompass the beta functions upon restricting to SNC geometry.}

{Observe that the Poisson equation $\langle S_+ \rangle = 0$ was found as a beta function of the SNC NR string. It thus does not suffice to take the NR limit of the NS-NS gravity action to retrieve the correct background dynamics of NR string theory. It shows that the full set of beta functions of non-relativistic string theory does not follow from an action principle. This is different from the analogous situation in relativistic string theory, where all the beta function constraints follow from a low-energy effective action.}

The emerging dilatation invariance that is present in the NR string worldsheet actions and that we have shown to be preserved in the NR limit of NS-NS gravity, hints that DSNC geometry is a natural target space geometry of NR string theory. In this regard, it is highly suggestive that the zero torsion beta function calculation of \cite{Gomis:2019zyu,Yan:2019xsf} can be generalized to worldsheet actions for strings in DSNC backgrounds.\footnote{In this regard, it is useful to mention that, upon quantization, a term proportional to $\lambda \bar{\lambda}$ is generically generated in the string worldsheet action \eqref{eq:nraction}. This term turns the string sigma model into a relativistic one. In \cite{Gomis:2019zyu,Yan:2019xsf}, the appearance of this term was prevented by restricting to zero torsion. Since this condition is tied to the existence of the $Z_A$ symmetry, mentioned at the end of section \ref{sec:NRstringaction}, one can then argue that requiring preservation of this symmetry in the quantum theory protects one from changing the theory to a relativistic one. Since here we do not require zero torsion, we also do not require the presence of the $Z_A$ symmetry. As a consequence, the quantum theory will then contain a $\lambda \bar{\lambda}$ term. However, it was pointed out to us by Ziqi Yan that the coefficient of this term vanishes upon imposing the non-linear constraint $\langle S_- \rangle = 0$, so that the quantum theory can still be regarded as a non-relativistic one. We thank Ziqi Yan for discussions on this point.} It would be interesting to calculate these beta functions and compare them with the NR limit of the equations of motion of NS-NS gravity, obtained in this paper.

\section{Conclusions}

In this paper we showed that a NR limit of the NS-NS gravity action can be defined which is based upon a crucial cancellation  of the quadratic divergences originating from the spin-connection squared terms  in the Einstein-Hilbert term with those arising from the kinetic term  of the KR 2-form field. These cancellations are the target space version of a similar collaboration of divergences that takes place when defining the NR limit of the relativistic string sigma model. They enable us to define a finite NR NS-NS gravity action without imposing any geometric constraint such as the zero torsion constraint which we imposed in our earlier work. Both the NR string sigma model and the NR NS-NS gravity action exhibit an emergent local dilatation symmetry.\,\footnote{
This  scale invariance only works in the directions longitudinal to the string. This is reminiscent of the reduced conformal symmetry recently discussed in \cite{Karch:2020yuy}.}
This emergent  symmetry has the effect that taking the limit of the relativistic NS-NS gravity action produces a NR action that leads to one equation of motion less than the ones that one obtains by taking the NR limit of the relativistic NS-NS gravity equations of motion. This missing equation of motion is precisely the Poisson equation of the Newton potential. This is consistent with the fact that it is not known how to obtain an action for NC gravity including the Poisson equation by taking a limit of General Relativity. In this paper we only consider  NR limits. If one would consider expansions (without taking the limit that $c$ goes to infinity) and/or more symmetries
than (a string extension of) the Bargmann algebra with more fields than the standard formulation of NC gravity, there are other ways to construct actions for non-relativistic gravity, see, e.g.,
\cite{Bergshoeff:2016lwr, Hartong:2016yrf, Hansen:2018ofj, Hansen:2019vqf}.  Our results imply that the NR equations of motion form a reducible, indecomposable representation of the String Galilei symmetries: it is consistent to delete the Poisson equation and obtain the same representation of equations of motion as the one that follows from varying the NR NS-NS gravity action.

In our approach, no geometric constraints are imposed by hand, but instead, the allowed geometry follows by solving the equations of motion. One natural geometry that in particular solves the nonlinear constraint equation \eqref{eq:S-} is given by a dilatation-invariant extension of the zero torsion constraint \eqref{SNCconstraints} which we called Dilatation invariant  String Newton-Cartan (DSNC) geometry. The constraints defining this geometry are given in eq.~\eqref{TTS}.

It would be interesting to see whether the same equations of motion that we obtained in this work by taking a NR limit can be obtained by redoing the beta function calculations of \cite{Gomis:2019zyu,Yan:2019xsf} in the absence of the zero torsion constraint, which breaks the local dilatation invariance. We compared our  NR equations of motion  with the beta function calculations of \cite{Gomis:2019zyu,Yan:2019xsf} for zero torsion and found a one-to-one correspondence except that we have one equation more that does not correspond to any beta function. It is given by the nonlinear constraint \eqref{eq:S-}. Remarkably, this constraint occurs as the coefficient in front {a $\lambda\bar\lambda$ operator that is generated by quantum corrections in the sigma model effective action. Such a term would deform the theory towards relativistic string theory as shown in \cite{Yan:2019xsf}. Equation \eqref{eq:S-} thus provides an ad-hoc obstruction to generating such a term in the quantum effective action. It would be interesting to understand these constraints from symmetry arguments.\footnote{Z. Yan, private communication.}}

It is  of interest to also compare our results with the beta function calculations of \cite{Gallegos:2019icg}. The starting point of \cite{Gallegos:2019icg} is the relativistic Polyakov sigma model in a background geometry with a null isometry direction. The beta function calculations were performed by first rewriting the  string sigma model in terms of NC variables in one dimension lower. In the light of this paper, it seems plausible that the beta functions calculated in \cite{Gallegos:2019icg}, except for the Poisson equation,  can  be identified with the equations that follow from the null reduction of the relativistic NS-NS gravity action. Usually, this null reduction is only performed at the level of the equations of motion with the argument that otherwise one misses the equation of motion that follows from varying the Kaluza-Klein scalar, which by the null isometry direction is set to zero. But this is precisely the Poisson equation, which, as we discussed in this paper, is not expected to follow  from an action but does follow from a null reduction of the relativistic equations of motion. The picture that arises is that the double dimensional reduction of the NR NS-NS gravity action we constructed in this paper is precisely the same action that one obtains from a null reduction of the relativistic NS-NS gravity action.\,\footnote{Note that one can also perform a direct dimensional reduction of the NR NS-NS gravity action along a transversal spatial direction leading to a sector of NR string theory that does not follow from  a null reduction of the relativistic NS-NS gravity action.} Neither action gives rise to the Poisson equation of the Newton potential but for different reasons: local dilatation symmetry versus null isometry.
For more details about this picture from a target space point of view, see \cite{inpreparation}.

In the context of string theory, it is well-known that there exists a so-called dual formulation of the relativistic NS-NS action where the 2-form KR field has been replaced, via an on-shell duality relation, by a 6-form potential that  couples to an NS-NS 5-brane. The dual action requires a different limit, which is characterized by a 5-brane foliation instead of the string foliation we used in this work. Whereas the string limit leads to a decoupling of all states that are not critically charged under the KR field, the 5-brane limit leads to a different decoupling where all states that are not critically charged with respect to the 6-form potential are decoupled \cite{Gomis:2000bd}.
We can consider  a 5-brane limit of the relativistic NS-NS gravity action in the same way that we took the string limit of the same action in this paper. We checked that the same crucial cancellation of infinities as in the case of the string limit takes place but now between the spin-connection squared term and the kinetic term for the  6-form potential. It would be interesting to see whether one could map the string and 5-brane  actions into each other thereby establishing a NR duality relation that maps a  solution of the NR string action to a dual solution of the NR 5-brane action, possibly via dimensional reduction to six dimensions.

The expression for the NR NS-NS gravity action that we derived in this work by taking a NR limit seems identical to the action recently derived in \cite{Gallegos:2020egk} from a  Double Field Theory (DFT) point of view. Also there, the Poisson equation takes a special status. In fact, this is a general phenomenon as discussed in \cite{Cho:2019ofr}.  This relation between DFT and NR string theory was already advocated some time ago \cite{Ko:2015rha,Morand:2017fnv}.\footnote{For other recent work on the NR string theory in DFT, see \cite{Blair:2020gng, Park:2020ixf}.} It is based upon the observation  that in DFT it is natural to use a degenerate geometry with a null isometry. What is intriguing and what we learned in this work is that imposing a null isometry from one point of view is the same as taking a NR limit from a T-dual point of view. It suggests that one should perhaps also be able to understand the results of this paper by defining a proper NR limit of DFT itself.

The present work grew out of an effort to define the NR limit of heterotic supergravity. Now a proper understanding of the bosonic case has been  obtained we hope to tackle the supersymmetric case soon. One additional complication in the supersymmetric case is that the spin-connection now not only gives rise to a quadratic divergence in the action but also to a linear divergence in the supersymmetry rules. The quadratic divergence in the action could be cancelled by a similar quadratic divergence due to the kinetic term of the KR 2-form field. However, the linear divergence in the supersymmetry rule due to the spin-connection cannot be cancelled by a similar linear divergence due to the KR curvature because the two fields have opposite parity with respect to worldsheet reflections. What can happen and what actually does happen is that in the relativistic case the two fields occur in the combination of a torsionful spin-connection
\begin{equation}
\Omega_\mu{}^{\hat A \hat B}(E, \mathcal H)  = \Omega_\mu{}^{\hat A \hat B}(E) + \mathcal H_\mu{}^{\hat A \hat B}\,,
\end{equation}
which leads to a linear divergence that is proportional to a self-dual projection of the torsion components that define the DSNC geometry
given in eq.~\eqref{TTS}. In order to obtain finite supersymmetry rules, we have to set these self-dual projections to zero by hand:
\begin{equation}\label{TTSNC-}
\tau_{A'+}{}^- = \tau_{A'B'}{}^- =0\,.
\end{equation}
This leads to a so-called self-dual DSNC geometry that seems to play a role in the supersymmetric case. One  attractive feature of the self-dual DSNC geometric constraints \eqref{TTSNC-}, not shared by the full DSNC geometric constraints \eqref{TTS}, is that the constraint equations \eqref{TTSNC-} are invariant under supersymmetry. We hope to give more details about this interesting case soon.

The results of this  paper can be used as a starting point for exploring many generalizations of NR string theory. For instance, one could  investigate whether one can give a meaning to the NR limit of IIA and IIB supergravity and even M-theory. Having supersymmetry under control one could investigate the presence of  half-supersymmetric fundamental string and other brane solutions to the equations of motion \cite{inpreparation}. Last but not least, one could consider taking a Carroll limit of NS-NS gravity and investigate whether an action for Carroll NS-NS gravity can be defined. We hope to come back to this and many other extensions, generalizations and applications in the nearby future.

\section*{Acknowledgements}

We thank Ziqi Yan, Jaume Gomis, Quim Gomis, Jelle Hartong, Niels Obers, Gerben Oling, Umut G\"ursoy, Natale Zinnato, Domingo Gallegos and Toine Van Proeyen for useful comments and discussions. JR also thanks the organizers of the NL Zoom Meeting on Non-Lorentzian geometry for the opportunity to present parts of the results of this paper which has led to useful discussions and comments. JR especially acknowledges Jeong-Hyuck Park for a useful email exchange regarding the DFT approach. We would also like to thank Microsoft Skype, Zoom, and our internet providers who made it possible to collaborate across borders during the COVID-19 pandemic. The work of C\c S is part of the research programme of the Foundation for Fundamental Research on Matter (FOM), which is financially supported by the Netherlands Organisation for Science Research (NWO). The work of LR is supported by the FOM/NWO free program {\sl Scanning New Horizons}.

\appendix

 \section{Conventions}\label{sec:conventions}

\subsection{Index Conventions} \label{ssec:indexconventions}

 In this paper, ten-dimensional curved indices are denoted by lowercase Greek letters. Ten-dimensional flat indices are denoted by $\hat A$, where $\hat{A} = 0,\cdots,9$. The index $\hat{A}$ is split into a `longitudinal' index $A$, where $A=0,1$, and a `transversal' index $A^\prime$, where $A'=2,\cdots,9$. We adopt the `mostly plus' convention for the ten-dimensional Minkowski metric $\eta_{\hat{A}\hat{B}}$, i.e., $\eta_{\hat{A}\hat{B}} = \text{diag}(-1,1,\cdots,1)$. Longitudinal indices $A$ are freely raised and lowered with $\eta_{AB}$, while transversal indices $A^\prime$ are raised and lowered with a Kronecker delta $\delta_{A^\prime B^\prime}$. The ten-dimensional epsilon symbols $\epsilon_{\hat{A}_0 \cdots \hat{A}_9}$ and $\epsilon^{\hat{A}_0 \cdots \hat{A}_9}$ are normalized as $\epsilon_{01\cdots 9}=+1$ and $\epsilon^{01\cdots 9}=-1$. We also use longitudinal epsilon symbols $\epsilon_{AB}$ and $\epsilon^{AB}$ that are normalized as $\epsilon_{01}=+1$ and $\epsilon^{01}=-1$. 
 We then have the following useful identities:
 \begin{equation}
   \label{eq:epsids}
   \epsilon_{AC} \epsilon_{BD} = -\eta_{AB} \eta_{CD} + \eta_{AD} \eta_{BC} \,, \qquad \qquad \epsilon_A{}^C \epsilon_{CB} = \eta_{AB} \,.
 \end{equation}
Symmetrization and antisymmetrization is defined with weight one, e.g.,
\begin{align}
    A_{[\mu\nu]}=\frac12\lr A_{\mu\nu} - A_{\nu\mu}\rr\,,\hskip 1.5truecm A_{(\mu\nu)} = \frac12\lr A_{\mu\nu} + A_{\nu\mu}\rr\,.
\end{align}
We furthermore use curly parentheses to denote traceless symmetric parts, e.g.,
\begin{equation}
  \label{eq:tracelesssymm}
  S_{\{AB\}} = S_{(AB)} - \frac12 \eta_{AB} S_C{}^C \qquad \text{with } S_C{}^C = \eta^{CD} S_{CD} \,.
\end{equation}

\subsection{Lorentzian Geometry Conventions} \label{ssec:LorGeom}

We denote the Vielbein of ten-dimensional Lorentzian geometry by $ E_\mu{}^{\hat A}$ and its inverse by $ E_{\hat A}{}^\mu$:
\begin{align} \label{eq:invE}
    {E}_\mu{}^{\hat{A}} E_{\hat{B}}{}^\mu = \delta^{\hat{A}}{}_{\hat{B}}\,,
        \qquad
    {E}_\mu{}^{\hat{A}} E_{\hat{A}}{}^\nu = \delta_\mu^\nu\,.
\end{align}
The fields $E_\mu{}^{\hat A}$ and $E_{\hat A}{}^\mu$ transform as a one-form, resp. vector under general coordinate transformations and as follows under local SO$(1,9)$ Lorentz transformations with parameter $\Lambda^{\hat{A}\hat{B}} = -\Lambda^{\hat{B}\hat{A}}$:
\begin{align}\label{eq:LoLo}
    \delta  E_\mu{}^{\hat A} =  \Lambda^{\hat A}{}_{\hat B}  E_\mu{}^{\hat B}\,,\hskip 1.5truecm \delta  E_{\hat A}{}^\mu = \Lambda_{\hat A}{}^{\hat B} E_{\hat B}{}^{\mu}\,.
\end{align}
{Upon splitting the ten-dimensional frame indices into a longitudinal  and a transversal part, we find that
\begin{align}
    \delta\,E_\mu{}^A = \Lambda^A{}_B E_\mu{}^B + \Lambda^A{}_{A'} E_\mu{}^{A'}\,,\hskip 1.5truecm \delta\,E_\mu{}^{A'} = \Lambda^{A'}{}_{B'} E_\mu{}^{B'}  - \Lambda_A{}^{A'}E_\mu{}^{A}\,,
\end{align}
and similarly for the inverse Vielbein.}\\
We denote the torsionless spin connection of Lorentzian geometry by $\Omega_\mu{}^{\hat{A}\hat{B}}$. In this paper, we work in the second-order formalism and define $\Omega_\mu{}^{\hat{A}\hat{B}}$ as the solution of the following zero torsion constraint:
\begin{align}
    \mcR_{\mu\nu}(P^{\hat A}) = 2\,\partial_{[\mu}E_{\nu]}{}^{\hat A} - 2\,\Omega_{[\mu}{}^{\hat A\hat B}E_{\nu]\hat B} = 0\,.
\end{align}
Explicitly, one has
\begin{align}\label{eq:relSpinConn}
    \Omega_\mu{}^{\hat A\hat B} =  E_{\mu\hat C} E^{\hat A\hat B\hat C} - 2 E_\mu{}^{[\hat A\hat B]}\,,\qquad\mathrm{where}\qquad  E_{\mu\nu}{}^{\hat A} = \partial_{[\mu} E_{\nu]}{}^{\hat A}\,.
\end{align}
The spin connection then transforms under local Lorentz transformations as an SO$(1,9)$ connection:
\begin{align}
    \delta \Omega_\mu{}^{\hat A\hat B} = \partial_\mu \Lambda^{\hat A\hat B} - 2\,\Omega_\mu{}^{\hat C[\hat A}\Lambda^{\hat B]}{}_{\hat C} \,.
\end{align}
The covariant curvature 2-form, associated to $\Omega_\mu{}^{\hat{A}\hat{B}}$ is defined by:
\begin{align}\label{eq:curv2}
    \mathcal{R}_{\mu\nu}{}^{\hat A\hat B} = 2\,\partial_{[\mu}\Omega_{\nu]}{}^{\hat A\hat B}  + 2\,\Omega_{[\mu}{}^{\hat A\hat C}\Omega_{\nu]}{}^{\hat B}{}_{\hat C}\,.
\end{align}
These quantities are related to the Christoffel connection $\Gamma_{\mu\nu}{}^\rho$ and the Riemann tensor $\mathcal{R}_{\mu\nu}{}^\rho{}_\sigma$ as follows
\begin{align}
    \Gamma_{\mu\nu}{}^{\rho} =  E_{\hat A}{}^{\rho}\lr\partial_{(\mu} E_{\nu)}{}^{\hat A} - \Omega_{(\mu}{}^{\hat A\hat B} E_{\nu)\hat B}\rr\qquad\mathrm{and}\qquad {\mathcal{R}}^\rho{}_{\sigma\mu\nu} = - E^\rho{}_{\hat A} E_{\sigma\hat B}\mathcal{R}_{\mu\nu}{}^{\hat A\hat B}\,.
\end{align}
The Ricci tensor and scalar are then expressed in terms of the curvature 2-form $\mathcal{R}_{\mu\nu}{}^{\hat{A}\hat{B}}$ as:
\begin{align}\label{eq:Ricci}
    \mathcal R_{\mu\nu} = \mcR^\rho{}_{\mu\rho\nu}= - E^\rho{}_{\hat A}\mathcal{R}_{\rho\mu}{}^{\hat A\hat B} E_{\nu\hat B}\,\qquad\mathrm{and}\qquad \mathcal R = - E^\mu{}_{\hat A} E^\nu{}_{\hat B}\mathcal{R}_{\mu\nu}{}^{\hat A\hat B}\,.
\end{align}

\subsection{Conversion of Curved to Flat Indices} \label{ssec:indexconversions}

The curved indices on a tensor $X^{\mu_1 \cdots \mu_r}{}_{\mu_{r+1} \cdots \mu_p}$ in the relativistic theory are turned into flat ones, using the relativistic (inverse) Vielbein $E_\mu{}^{\hat{A}}$ ($E_{\hat{A}}{}^\mu$) in the usual fashion
\begin{equation}
  X^{\hat{A}_1 \cdots \hat{A}_r}{}_{\hat{A}_{r+1} \cdots \hat{A}_p} = E_{\mu_1}{}^{\hat{A}_1} \cdots E_{\mu_r}{}^{\hat{A}_r} E_{\hat{A}_{r+1}}{}^{\mu_{r+1}} \cdots E_{\hat{A}_p}{}^{\mu_p} X^{\mu_1 \cdots \mu_r}{}_{\mu_{r+1} \cdots \mu_p} \,.
\end{equation}
Note that the $\hat{A}$ index will often be split into a longitudinal index $A$ and a transversal index $A^\prime$. As an example, if $X_{\mu\nu}$ is a tensor in the relativistic theory, the quantities $X_{AB}$, $X_{A B^\prime}$, $X_{A^\prime B}$ and $X_{A^\prime B^\prime}$ are to be understood as
\begin{align}
  X_{AB} & = E_A{}^\mu E_B{}^\nu X_{\mu\nu} \,, \qquad X_{A B^\prime} = E_A{}^\mu E_{B^\prime}{}^\nu X_{\mu\nu} \,, \qquad X_{A^\prime B} = E_{A^\prime}{}^\mu E_B{}^\nu X_{\mu\nu} \,, \nonumber \\ X_{A^\prime B^\prime} &= E_{A^\prime}{}^\mu E_{B^\prime}{}^\nu X_{\mu\nu} \,.
\end{align}


The curved indices on tensors in the non-relativistic theory are turned into flat ones, using the longitudinal and transverse Vielbeine $\tau_\mu{}^A$, $e_\mu{}^{A^\prime}$.
For example, if $Y^\mu{}_\nu$ is a tensor in the non-relativistic theory, the objects $Y^A{}_B$, $Y^A{}_{A^\prime}$, $Y^{A^\prime}{}_A$ and $Y^{A^\prime}{}_{B^\prime}$ are defined as
\begin{align}
  Y^A{}_B &= \tau_\mu{}^A \tau_B{}^\nu Y^\mu{}_\nu \,, \qquad Y^A{}_{A^\prime} = \tau_\mu{}^A e_{A^\prime}{}^\nu Y^\mu{}_\nu \,, \qquad Y^{A^\prime}{}_A = e_\mu{}^{A^\prime} \tau_A{}^{\nu} Y^\mu{}_\nu \,, \nonumber \\ Y^{A^\prime}{}_{B^\prime} &= e_\mu{}^{A^\prime} e_{B^\prime}{}^\nu Y^\mu{}_\nu \,.
\end{align}

\section{{Torsional String Newton Cartan Geometry}}\label{sec:stringGalileigeometry}

In this section, we give details on the non-Lorentzian geometry that appears in the NR limit of NS-NS gravity, discussed in this paper. We refer to this geometry as `torsional string Newton Cartan geometry' (TSNC). The basic geometric fields are the longitudinal Vielbein $\tau_\mu{}^A$ ($A=0,1$), the transverse Vielbein $e_\mu{}^{A^\prime}$ ($A^\prime = 2,\cdots,9$), the KR field $b_{\mu\nu}$ and the dilaton $\phi$. These fields transform under local String Galilei symmetries (longitudinal SO$(1,1)$ Lorentz transformations, transverse SO$(8)$ rotations and Galilean boosts) and local dilatations, according to:
\begin{align} \label{eq:trafosSGG}
  \delta\tau_\mu{}^{A} &= \lambda_M\,\epsilon^A{}_B\tau_\mu{}^B + \lambda_D \tau_\mu{}^A \,, & \delta e_\mu{}^{A'} &= \lambda^{A'}{}_{B'}e_\mu{}^{B'} - \lambda_A{}^{A'}\tau_\mu{}^{A}\,, \nonumber \\
  \delta b_{\mu\nu} &= - 2\,\epsilon_{AB}\lambda^A{}_{A^\prime}\,\tau_{[\mu}{}^{B}e_{\nu]}{}^{A^\prime} \,, & \delta \phi &= \lambda_D \,,
\end{align}
where $\lambda_M$, $\lambda^{A^\prime B^\prime}$, $\lambda^{A A^\prime}$, $\lambda_D$ are the parameters of SO$(1,1)$, SO$(8)$, Galilean boosts and dilatations resp. The KR field is also subjected to an abelian two-form symmetry:
\begin{equation}
  \delta b_{\mu\nu} = 2\,\partial_{[\mu} \theta_{\nu]} \,.
\end{equation}
Projective inverses $\tau_A{}^\mu$ and $e_{A^\prime}{}^\mu$ are introduced via \eqref{eq:invvielbeine}. They transform as
\begin{align}
  \delta\tau_A{}^\mu = \lambda_M\,\epsilon_A{}^{B}\tau_B{}^{\mu} + \lambda_A{}^{A'}e_{A'}{}^{\mu} - \lambda_D \tau_A{}^\mu \,, \qquad \qquad \delta e_{A'}{}^\mu = \lambda_{A'}{}^{B^\prime} e_{B^\prime}{}^\mu \,.
\end{align}
Note that the KR field transforms non-trivially to the longitudinal and transverse Vielbeine under Galilean boosts, while the dilaton acquires a shift under dilatations. For this reason, we treat $b_{\mu\nu}$ and $\phi$ as part of the geometric data. This should be contrasted with relativistic string theory/NS-NS gravity, in which the KR field and dilaton are treated as matter fields, instead of geometric fields.

In the following, we will discuss how these fields can be used to define connections and curvatures for local $\mathrm{SO}(1,1) \times \mathrm{SO}(8)$ transformations, Galilean boosts and dilatations. In Appendix \ref{sec:SGGinlimit} we will then describe how the geometric structures, described here, appear in the limit of the action and equations of motion of NS-NS gravity.

\subsection{String Galilei and Dilatation Connections}

Here, we introduce String Galilei spin connections $\omega_\mu$, $\omega_\mu{}^{A A^\prime}$ and $\omega_\mu{}^{A^\prime B^\prime}$ for SO$(1,1)$ Lorentz transformations, SO$(8)$ rotations and Galilean boosts resp., as well as a dilatation connection $b_\mu$. In analogy to the spin connection $\Omega_\mu{}^{\hat{A}\hat{B}}$ of Lorentzian geometry, we will define these connections as expressions that depend on the geometric data $\tau_\mu{}^A$, $e_\mu{}^{A^\prime}$, $b_{\mu\nu}$ and $\phi$, with correct transformation properties. In particular, we seek to define $\omega_\mu$, $\omega_\mu{}^{A A^\prime}$, $\omega_\mu{}^{A^\prime B^\prime}$ and $b_\mu$ as dependent expressions that transform under String Galilei transformations and dilatations as
\begin{subequations}\label{eq:spinconntranslin}
    \begin{align}
    &\delta b_\mu = \partial_\mu\lambda_D+\cdots\,,&&\delta\omega_\mu=\partial_\mu\lambda_M + \cdots\,,\\
    &\delta\omega_\mu{}^{A A^\prime} = \partial_\mu\lambda^{A A^\prime}+\cdots\,,&&\delta\omega_\mu{}^{A^\prime B^\prime} = \partial_\mu\lambda^{A^\prime B^\prime}+\cdots\,,
\end{align}
\end{subequations}
where the ellipses denote terms that do not involve derivatives of a parameter. To do this, we consider the following `covariant' quantities
\begin{subequations}
\label{eq:cartan1ingredients}
\begin{align}
    \nabla_\mu\phi &\equiv \partial_\mu\phi - b_\mu\,,\\
    \rmR_{\mu\nu}(H^A) &\equiv 2\,\tau_{\mu\nu}{}^A - 2\,\big(\epsilon^A{}_B\,\omega_{[\mu} + \delta^A{}_B\,b_{[\mu}\big)\,\tau_{\nu]}{}^B\,,\\
    \rmR_{\mu\nu}(P^{A'}) &\equiv 2\,e_{\mu\nu}{}^{A'} -2\,\omega_{[\mu}{}^{A'B'}\,e_{\nu]B'} + 2\,\omega_{[\mu}{}^{A A^\prime}\,\tau_{\nu]A}\,,\\
    H_{\mu\nu\rho} &\equiv h_{\mu\nu\rho} + 6\,\epsilon_{AB}\,\omega_{[\mu}{}^{AB'}\tau_{\nu}{}^B\,e_{\rho]B'}\,.
\end{align}
\end{subequations}
where we have defined
\begin{align} \label{eq:deftaueh}
    \tau_{\mu\nu}{}^A = \partial_{[\mu}\tau_{\nu]}{}^A\,,\qquad e_{\mu\nu}{}^{A'} = \partial_{[\mu}e_{\nu]}{}^{A'}\,, \qquad h_{\mu\nu\rho}=3\,\partial_{[\mu}b_{\nu\rho]} \,.
\end{align}
The quantities defined in \eqref{eq:cartan1ingredients} are covariant in the sense that they transform without derivatives of a parameter, if the transformation rules \eqref{eq:spinconntranslin} hold.

Similarly to how one defines the relativistic spin connection \eqref{eq:relSpinConn}, we can then try to express the String Galilei spin connections and the dilatation connection as dependent fields that solve conventional constraints,\footnote{By conventional constraints we mean constraints that reduce the number of independent fields in a theory, i.e., constraints that contain certain fields algebraically and that can be used to solve those fields in terms of other fields.} that are obtained by putting certain components of the covariant quantities \eqref{eq:cartan1ingredients} equal to zero. Note that we should only constrain those parts of \eqref{eq:cartan1ingredients} that contain components of $\omega_\mu$, $\omega_\mu{}^{A A^\prime}$, $\omega_\mu{}^{A^\prime B^\prime}$ and $b_\mu$. In particular, the following components of \eqref{eq:cartan1ingredients}
\begin{align}
    \rmR_{A'B'}(H_C) = 2\,\tau_{A'B'C}\,,\qquad \rmR_{A'\{A}(H_{B\}}) = 2\,\tau_{A'\{AB\}}\,,\qquad H_{A'B'C'} = h_{A'B'C'}\,,
\end{align}
are independent of the String Galilei spin connections and $b_\mu$ and are not set to zero as conventional constraints. Note that $\rmR_{A'B'}(H_C)$ and $\rmR_{A'\{A}(H_{B\}})$ contain information about the intrinsic torsion of the geometry \cite{Figueroa-OFarrill:2020gpr}. For the remaining components of \eqref{eq:cartan1ingredients}, we then adopt the following constraints
\begin{subequations}\label{eq:cartan1}
\begin{align}
    &\nabla_A\phi = 0\,,&&\eta^{AB}\rmR_{A'A}(H_B)=0\,,\label{eq:cartan1dil}\\
    &\epsilon^{AB}\,\rmR_{A'A}(H_B)=0\,,&& \epsilon^{AB}\rmR_{AB}(H_C)=0\,,\label{eq:cartan1long}\\
    &\rmR_{\mu\nu}(P^{A'})=0\,,\label{eq:cartan1trans}\\
    &H_{AA'B'}=0\,,&&H_{ABA'}=0\label{eq:cartan1boost}\,.
\end{align}
\end{subequations}
These can be viewed as $444$ algebraic equations for the $460$ components of $b_\mu$, $\omega_\mu$, $\omega_\mu{}^{AA'}$, $\omega_\mu{}^{A'B'}$. These equations are thus not able to determine all components of the String Galilei spin connections and dilatation field\footnote{A similar phenomenon has been encountered in \cite{Bergshoeff:2019pij}}. The 16 undetermined components reside in $\omega_{\{AB\}}{}^{A^\prime}$ and will in the following be denoted by $W_{\mu A}{}^{A'} = \tau_\mu{}^B\omega_{\{AB\}}{}^{A'}$. We can then express the most general solution of the conventional constraints \eqref{eq:cartan1} by
\begin{subequations}\label{eq:galspinconn}
\begin{align}
    b_\mu &= e_\mu{}^{A'}\,\tau_{A'A}{}^A +\tau_\mu{}^A\partial_A\phi\,,\\
    \omega_\mu &= \big(\,\tau_\mu{}^{AB}-\frac12\,\tau_\mu{}^C\tau^{AB}{}_C \big)\epsilon_{AB} - \tau_\mu{}^A\,\epsilon_{AB}\partial^B\phi\,,\\
 \omega_\mu{}^{AA'} &=   -e_\mu{}^{AA'}+e_{\mu B'}e^{AA'B'} + \frac12\,\epsilon^A{}_B\,h_\mu{}^{BA'} + W_\mu{}^{AA'}\,,\\
    \omega_\mu{}^{A'B'} &= -2\, e_{\mu}{}^{[A'B']}+e_{\mu C'}e^{A'B'C'} - \frac12\,\tau_\mu{}^A\,\epsilon_{AB}\,h^{BA'B'}\,.
\end{align}
\end{subequations}
The transformation rules of \eqref{eq:galspinconn} under String Galilei symmetries and dilatations can be obtained by requiring that the set of constraints \eqref{eq:cartan1} does not transform under these symmetries. Writing down these requirements, using \eqref{eq:trafosSGG}, one obtains equations for $\delta b_\mu$, $\delta \omega_\mu$, $\delta \omega_\mu{}^{A A^\prime}$ and $\delta \omega_\mu{}^{A^\prime B^\prime}$ that can be solved to give\footnote{The undetermined components in the boost spin connection transform as follows
\begin{align}
        \delta W_\mu{}^{AA'} &= \tau_\mu{}_B\big(\nabla^{\{A}\lambda^{B\}A'} - \omega^{B'A'\{A}\lambda^{B\}}{}_{B'}\big)\notag\\
        &\quad+\lambda_M\,\epsilon^A{}_B\,W_\mu{}^{BA'} + \lambda^{A'}{}_{B'}W_\mu{}^{AB'} - \lambda_D\,W_\mu{}^{AA'}\,.
\end{align}
}
{\begin{subequations}\label{eq:deltaomega1,18}
    \begin{align}
        \delta\,b_\mu &= \partial_\mu\lambda_D\,,\hskip 2truecm \delta\,\omega_\mu = \partial_\mu\lambda_M\,,\\
        \delta\,\omega_\mu{}^{A'B'} &= \partial_\mu\lambda^{A'B'} - 2\,\omega_\mu{}^{C'[A'}\lambda^{B']}{}_{C'}\,,\\
        \delta\,\omega_\mu{}^{AA'}& = \lambda_M\,\epsilon^A{}_B\,\omega_\mu{}^{BA'} + \lambda^{A'}{}_{B'}\omega_\mu{}^{AB'} - \lambda_D\,\omega_\mu{}^{AA'}\,,
    \end{align}
\end{subequations}}
for the transformations under $\mathrm{SO}(1,1)\times\mathrm{SO}(8)$ and dilatations and
{\begin{subequations}\label{eq:deltaomegaboost}
    \begin{align}
    \delta\,b_\mu &=
                    e_\mu{}^{A'}\,\tau_{A'B'B}\lambda^{BB'} + \tau_\mu{}^A\lambda_{AA'}\nabla^{A'}\phi\,,\\
    \delta\,\omega_\mu &=
                    -e_\mu{}^{A'}\epsilon^{AB}\tau_{A'B'A}\lambda_B{}^{B'} - 2\,\tau_\mu{}^A\big(\epsilon^{BC}\lambda_{B}{}^{B'}\tau_{B'\{AC\}} + \frac12\epsilon_{AB}\lambda^{BB'}\nabla _{B'}\phi\big)\,,\\
    \delta\,\omega_\mu{}^{AA'} &=
                    \nabla_\mu\lambda^{AA'} + 2\,e_\mu{}^{B'}\big(\lambda_{BB'}\tau^{A'\{AB\}} + \frac14\,\epsilon^{AB}\lambda_{BC'}\,h^{A'B'C'}\big)\,,\\
    \delta\,\omega_\mu{}^{A'B'} &=
                    4\,\tau_\mu{}^A\big(\lambda^{B[A'}\tau^{B']}{}_{\{AB\}} - \frac18\,\epsilon_{AB}\lambda^B{}_{C'}\,h^{A'B'C'}\big)\notag\\
                    &\quad - e_\mu{}^{C'}\big(\lambda_{CC'}\tau^{A'B'C} - 2\,\lambda_C{}^{[A'}\tau^{B']C'C}\big) \,,
\end{align}
\end{subequations}}
under Galilean boosts, where we have defined
\begin{align}
    \nabla_\mu\lambda^{AA'} = \partial_\mu\lambda^{AA'} - \omega_\mu\,\epsilon^A{}_B\lambda^{BA'} - \omega_\mu{}^{A'B'}\lambda^A{}_{B'} + b_\mu\,\lambda^{AA'}\,.
\end{align}

\subsection{Affine Connection}

Using the String Galilei spin connections and dilatation connection, we can then also introduce an affine, metric compatible connection $\Gamma^\rho_{\mu\nu}$, by imposing the following Vielbein postulates
\begin{subequations}\label{eq:metcomp}
    \begin{align}
    \nabla_\mu\tau_\nu{}^A &= \partial_\mu\tau_\nu{}^A - \omega_\mu\,\epsilon^{AB}\tau_{\nu B} - b_\mu\,\tau_\nu{}^A - \Gamma_{\mu\nu}^\rho\tau_\rho{}^A = 0\,,\\
    \nabla_\mu e_\nu{}^{A'} &= \partial_\mu e_\nu{}^{A'} - \omega_\mu{}^{A'B'} e_{\nu B'} + \omega_\mu{}^{AA'}\tau_{\nu A} - \Gamma_{\mu\nu}^\rho e_\rho{}^{A'}=0\,.
\end{align}
\end{subequations}
{This connection has intrinsic torsion
\begin{align}
    T_{\mu\nu}^\rho = 2\,\Gamma_{[\mu\nu]}^\rho = R_{\mu\nu}(H^A)\tau_A{}^\rho \,.
\end{align}
As a corollary of \eqref{eq:metcomp}, we derive the following identities:
\begin{subequations}
    \begin{align}
    \partial_\mu\big(e\,e_{A'}{}^\mu\big) &= e\,e_{B'}{}^\mu\omega_{\mu A'}{}^{B'} + 2\,e\,b_{A'}\,,\label{eq:partorpe}\\
    \partial_\mu\big(e\,\tau_A{}^\mu\big) &= e\,\tau_B{}^\mu\big(\epsilon_A{}^B\,\omega_\mu + \delta_A{}^B\,b_\mu\big) + e\,e_{A'}{}^\mu\omega_{\mu A}{}^{A'}\,,\label{eq:partorptau}
    \end{align}
\end{subequations}}
which are used to derive the action \eqref{eq:NRaction2} and the equations of motion \eqref{eq:nrEOM}.

\subsection{Curvatures}
Using the transformation rules for the spin connections \eqref{eq:deltaomega1,18} and \eqref{eq:deltaomegaboost} we can define covariant curvature $2-$forms
\begin{subequations}\label{eq:curv2forms}
\begin{align}
    \rmR_{\mu\nu}(D)         &=
                    2\,\partial_{[\mu}b_{\nu]} {+ 2\,e_{[\mu}{}^{A'}\omega_{\nu]}{}^{BB'}}\,\tau_{A'B'B} + 2\,\tau_{[\mu}{}^A\,\omega_{\nu]A}{}^{A'}\,\nabla _{A'}\phi\,,\\
    \rmR_{\mu\nu}(M)         &=
                    2\,\partial_{[\mu}\omega_{\nu]}+ 2\,\epsilon_{AB}\,e_{[\mu}{}^{A'}\omega_{\nu]}{}^{AB'}\,\tau_{A'B'}{}^{B}\notag\\
                    &\quad - 4\,\tau_{[\mu}{}^A\omega_{\nu]}{}^{BB'}\,\epsilon_B{}^C\tau_{B'\{AC\}} + 2\,\epsilon_{AB}\,\tau_{[\mu}{}^A\omega_{\nu]}{}^{BB'}\,\nabla _{B'}\phi\,,\\
    \rmR_{\mu\nu}(G)^{AA'}   &=
                    2\,\partial_{[\mu}\omega_{\nu]}{}^{AA'}-2\,\epsilon^A{}_B\,\omega_{[\mu}{}\omega_{\nu]}{}^{BA'} - 2\,\omega_{[\mu}{}^{A'B'}\omega_{\nu]}{}^A{}_{B'} + 2\,b_{[\mu}{} \omega_{\nu]}{}^{AA'}\notag\\
                    &\quad - 4\,e_{[\mu}{}^{B'}\big(\omega_{\nu]B}{}^{[A'}\tau^{B']\{AB\}} -\frac14\,\epsilon^{AB}\,\omega_{\nu]BC'}\,h^{A'B'C'}\big) \,,\\
    \rmR_{\mu\nu}(J)^{A'B'}  &=
                    2\,\partial_{[\mu}\omega_{\nu]}{}^{A'B'} + 2\,\omega_{[\mu}{}^{A'C'}\omega_{\nu]}{}^{B'}{}_{C'}\notag\\
                    &\quad +2\,e_{[\mu}{}^{C'}\big( 2\,\omega_{\nu]}{}^{C[A'}\tau^{B']}{}_{C'C}{-\omega_{\nu]CC'}\tau^{A'B'C}}\big) \notag\\
                    &\quad+8\,\tau_{[\mu}{}^A\big(\omega_{\nu]}{}^{B[A'}\tau^{B']}{}_{\{AB\}}-\frac18\,\epsilon_{A}{}^B\,\omega_{\nu]BC'}h^{A'B'C'}\big)\,.
\end{align}
\end{subequations}
The Ricci scalar built from the curvature of transverse rotations reads
\begin{align}
    \rmR(J) &= -\rmR_{A'B'}(J)^{A'B'} \notag\\
            &= -2\,e_{A'}{}^\mu e_{B'}{}^\nu\big(\partial_{[\mu}\omega_{\nu]}{}^{A'B'} + \omega_{[\mu}{}^{A'C'}\omega_{\nu]}{}^{B'}{}_{C'}\big)(e_{A},e_{B'}) {-4\,\omega_{A'BB'}\tau^{A'B'B}}.\label{eq:transRicci}
\end{align}
Substituting the conventional constraints \eqref{eq:cartan1} into the Bianchi identities, one can derive the following relations:
\begin{subequations}\label{eq:Bianchi-components}
    \begin{align}
    \rmR_{[B'C'}(J)^{A'}{}_{D']} &= 0\,&&\longrightarrow& \rmR_{C'[A'}(J)_{B']}{}^{C'}&=0\,,\label{bian:Jsym}\\
    2\,\rmR_{[B'}{}^A(J)^{A'}{}_{C']} &= {- \rmR_{B'C'}(G)^{AA'}}\,&&\longrightarrow&\rmR_{C'A}(J)_{A'}{}^{C'}&=\rmR_{C'A'}(G)_A{}^{C'}\,,\\
    \nabla_A\,\tau_{A'B'}{}^A &= -\rmR_{A'B'}(D) &&\longrightarrow& \nabla_A\,\tau_{A'B'}{}^A &= 2\,\nabla_{[A'}\nabla_{B']}\phi\,.\label{bian:tauA'B'}
\end{align}
\end{subequations}

\section{Details on the NR Limit of NS-NS Gravity} \label{sec:SGGinlimit}

Here, we provide details on how the NR limit of NS-NS gravity is taken in section \ref{sec:NRNSNS}. In particular, we give the results of expanding the geometric objects of Lorentzian geometry of \ref{ssec:LorGeom} in powers of $c^{-2}$, after applying the redefinitions \eqref{eq:R=NR}. We also show how these results can be rewritten in terms of geometric quantities of TSNC geometry, discussed in the previous Appendix \ref{sec:stringGalileigeometry}. We refer to Appendices \ref{sec:conventions} and \ref{sec:stringGalileigeometry} for the notation used in this section.

Let us start by applying the redefinitions \eqref{eq:R=NR} to the relativistic spin connection \eqref{eq:relSpinConn} and expanding the result into leading and subleading orders of powers of $c^{-2}$. This leads to
\begin{subequations}
\label{eq:screscale}
\begin{align}
\Omega_{\mu}&=\accentset{(0)}{\omega}_{\mu}+\frac{1}{c^2}~\accentset{(-2)}{\omega}_{\mu}\,,\\
\Omega_{\mu}{}^{AB'}&=c\ \accentset{(1)}{\omega}_{\mu}{}^{AB'}+\frac{1}{c}~\accentset{(-1)}{\omega}_{\mu}{}^{AB'}{=-\Omega_\mu{}^{B'A}}\,,\\
\Omega_{\mu}{}^{A'B'}&=c^{2}\, \accentset{(2)}{\omega}_{\mu}{}^{A'B'}+
\accentset{(0)}{\omega}_{\mu}{}^{A'B'}\,,
\end{align}
\end{subequations}
where we have set $\Omega_\mu{}^{AB}=\Omega_\mu\,\epsilon^{AB}$ and we explicitly have
\begin{subequations}\label{eq:leading}
\begin{align}
&\accentset{(0)}{\omega}_{\mu}=\epsilon_{AB}\big(e_{\mu C'}{\tau}^{C'AB}-\tau_{\mu C}{\tau}^{ABC}\big)\,,&& \accentset{(-2)}{\omega}_{\mu}=-\frac12\,\epsilon_{AB} e_{\mu C'}{e}^{ABC'}\,,\\
&\accentset{(1)}{\omega}_\mu{}^{AA'}=  e_{\mu B'}\tau^{B'A'A}-2\,\tau_{\mu B}\tau^{A'(BA)}\,, &&\accentset{(-1)}{\omega}_\mu{}^{AA'}= 2\,e_{\mu B'}{e}^{A(A'B')}-\tau_{\mu B} e^{BAA'}\,,\\
&\accentset{(2)}{\omega}_\mu{}^{A'B'}= \tau_{\mu C}{\tau}^{A'B'C}\,, && \accentset{(0)}{\omega}_\mu{}^{A'B'}=e_{\mu C'}{e}^{A'B'C'}-2\, {e}_{\mu}{}^{[A'B']} \,.
\end{align}
\end{subequations}
It is useful to note that
\begin{align}
    &\accentset{(1)}{\omega}{}^{ABA'}=-2\,\tau^{A'(AB)}\,,
    &&\accentset{(1)}{\omega}_C{}^{CC'}=-2\,b^{C'}\,,\notag\\
    &\accentset{(1)}{\omega}{}^{B'AA'}=\tau^{B'A'A}\,,
    &&\accentset{(1)}{\omega}_{A'}{}^{AA'}=0\,,\\
    &\accentset{(2)}{\omega}{}^{AB'C'}=\tau^{B'C'A}\,,&&\accentset{(2)}{\omega}{}^{A'B'C'}=0\,.&&\notag
\end{align}
None of the expressions in \eqref{eq:leading} correspond to String Galilei spin connections as they stand; rather $\accentset{(0)}{\omega}_\mu$, $\accentset{(-1)}{\omega}_\mu{}^{AA'}$ and $\accentset{(0)}{\omega}_\mu{}^{A'B'}$ are related to the String Galilei spin connections $\omega_\mu$, $\omega_\mu{}^{A A^\prime}$ and $\omega_\mu{}^{A^\prime B^\prime}$ via
\begin{subequations}\label{eq:limitvsgeometry}
    \begin{align}
    \accentset{(0)}{\omega}_\mu &= \omega_\mu + \tau_\mu{}^A\epsilon_{AB}\partial^B\phi\,,\\
    \accentset{(-1)}{\omega}_\mu{}^{AA'} &= \omega_\mu{}^{AA'} - \frac12\,\epsilon^A{}_B\,h_\mu{}^{BA'} - W_\mu{}^{AA'}\,,\\
    \accentset{(0)}{\omega}_\mu{}^{A'B'} &= \omega_\mu{}^{A'B'} + \frac12\,\tau_\mu{}^A\,\epsilon_{AB}\,h^{BA'B'}\,.
\end{align}
\end{subequations}
Expanding $\mathcal{H}_{\mu\nu\rho}= 3\,\partial_{[\mu} B_{\nu\rho]}$ in powers of $c^{-2}$, using \eqref{eq:R=NR}, we have
\begin{align}\label{eq:Hexpans}
    &\mathcal{H}_{\mu\nu\rho} = c^2\,\accentset{(2)}{\mathcal{H}}_{\mu\nu\rho} + \accentset{(0)}{\mathcal{H}}_{\mu\nu\rho}\,,\\
    \mathrm{with}\qquad &\accentset{(2)}{\mathcal{H}}_{\mu\nu\rho} = 6\,\epsilon_{AB}\tau_{[\mu}{}^A\tau_{\nu\rho]}{}^B\,,\qquad \accentset{(0)}{\mathcal{H}}_{\mu\nu\rho}=h_{\mu\nu\rho} = 3\,\partial_{[\mu}b_{\nu\rho]}\,,\notag\\
    &\accentset{(2)}{\mathcal{H}}_{AA'B'} = 2\,\epsilon_{AB}\tau_{A'B'}{}^B\,,\qquad \accentset{(2)}{\mathcal{H}}_{ABC'} = -2\,\epsilon_{AB}b_{C'}\,,\notag\\
    &\accentset{(2)}{\mathcal{H}}_{A'B'C'} = 0\,,\hskip 1.1truecm \mathrm{and}\qquad \accentset{(2)}{\mathcal{H}}_{ABC} = 0\,.\notag
\end{align}
Similarly, we can also expand the Ricci scalar \eqref{eq:Ricci} as
\begin{align}
    \mcR &= c^2\accentset{(2)}{\mcR} + \accentset{(0)}{\mcR} + \mathcal{O}(c^{-2})\,,\\
    \mathrm{where}:\,\,\accentset{(2)}{\mcR} &= -\eta_{AB}\tau_{A'B'}{}^A\tau^{A'B'B}\,,\label{eq:ricci2}\\
    \accentset{(0)}{\mcR} &= -\accentset{(0)}{\rmR}_{A'B'}{}^{A'B'}-4\,e_{A'}{}^{\mu}\left(\partial_\mu b^{A^\prime} - \accentset{(0)}{\omega}_\mu{}^{A'B'}b_{B'} +\frac32\,b_{\mu}b^{A'}\right) - 4\,\tau_{A'\{AB\}}\tau^{A'\{AB\}}\,,\label{expansion:R0}
\end{align}
where $\accentset{(0)}{\rmR}_{A'B'}{}^{A'B'} = 2\,e_{A'}{}^\mu e_{B'}{}^\nu(\partial_{[\mu}\accentset{(0)}{\omega}_{\nu]}{}^{A'B'} + \accentset{(0)}{\omega}_{[\mu}{}^{C'A'}\accentset{(0)}{\omega}_{\nu]}{}^{B'}{}_{C'})$\,.

The quantity $\accentset{(0)}{R}$ can be expressed in terms of TSNC geometric variables, using the relations \eqref{eq:limitvsgeometry}
\begin{align}\label{eq:R0toRJ}
    \accentset{(0)}{\mcR} = \rmR(J) - 4\,\mathcal D_{A'}b^{A'} - 6\,b_{A'}b^{A'} {+2}\,\omega_{A'BB'}\tau^{A'B'B}- 4\,\tau_{A'\{AB\}}\tau^{A'\{AB\}}\,,
\end{align}
where the derivative $\mathcal D_\mu$ is defined in \eqref{eq:derbexpl}.


\providecommand{\href}[2]{#2}\begingroup\raggedright\endgroup

\end{document}